\documentclass[letterpaper]{revtex4-2}

\usepackage[bitstream-charter]{mathdesign}
\DeclareSymbolFont{usualmathcal}{OMS}{cmsy}{m}{n}
\DeclareSymbolFontAlphabet{\mathcal}{usualmathcal}

\usepackage{graphicx}
\usepackage[export]{adjustbox}
\usepackage{amsmath}
\usepackage{amsfonts}
 \usepackage{amssymb}
\usepackage{mhchem}
\usepackage{stmaryrd}
\usepackage{wrapfig}

\begin{document}
\title{Statistical physics, Bayesian inference and neural information processing}

\author{Erin Grant}
\affiliation{Gatsby Computational Neuroscience Unit \& Sainsbury Wellcome Centre, UCL, London, UK}
\author{Sandra B. Nestler}
\affiliation{Institute of Neuroscience and Medicine (INM-6) and Institute for Advanced Simulation (IAS-6) and JARA-Institute Brain Structure-Function Relationships (INM-10),
Jülich Research Centre, Jülich, Germany}
\affiliation{Department of Physics, Faculty 1, RWTH Aachen University, Aachen, Germany}
\author{Berfin \c{S}im\c{s}ek}
\affiliation{Chair of Statistical Field Theory (CSFT) \& Laboratory of Computational Neuroscience (LCN), École Polytechnique Fédérale de Lausanne}
\author{Sara A. Solla}
\affiliation{Department of Neuroscience, Northwestern University, Chicago, IL 60611, USA}
\affiliation{Department of Physics and Astronomy, Northwestern University, Evanston, IL 60208, USA}


\date{July 4-29, 2022 \\
 Statistical Physics of Machine Learning Summer School \\
 Les Houches}
\maketitle
\newpage

\vspace{10pt}
\noindent\rule{\textwidth}{1pt}
\tableofcontents
\newpage
\noindent\rule{\textwidth}{1pt}
\vspace{10pt}

\section{\textsc{Lecture 1:} Statistical physics, Bayesian inference, and neural information processing}
\graphicspath{ {./Les_Houches_0722_-_Lecture_1/images/} }

The brain interprets the world; an embodied brain causes actions that change the world. 
The brain interacts with an external dynamical system, the world, that follows causal relationships, as in $\vec{y}$ follows $\vec{x}$. 
\begin{figure*}[htb]
\centering\includegraphics[max width=50pt]{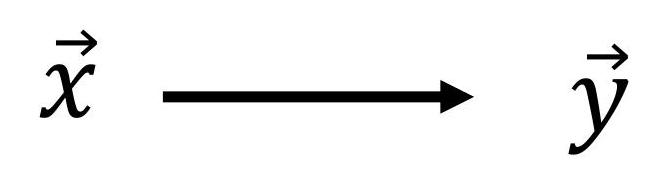}
\end{figure*}
However, the brain receives the same input $\vec{x}$, processes it, and acts in a manner that affects the output. 
\begin{figure}[htb]
\centering\includegraphics[max width=200pt]{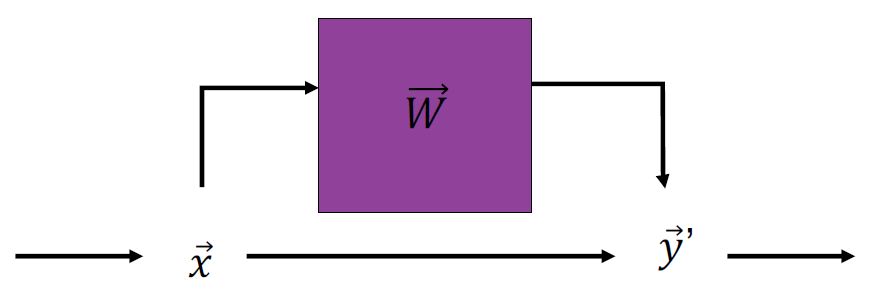}
\caption{The brain as part of the world: receiving inputs from the world and producing outputs that change the state of the world.} 
\end{figure}
This interactive process between the brain and the world can be framed in terms of input-output maps between high-dimensional state vectors 
$\vec{x}=\left\{x_{1}, x_{2}, \ldots, x_{N}\right\} \rightarrow \vec{y}=\left\{y_{1}, y_{2}, \ldots, y_{R}\right\}$.
The output of the system is a function of its input, 
\begin{equation}
\vec{y}=f(\vec{x})~.
\end{equation}

\subsection{Learning to model an input-output map}
We are particularly interested in the case in which the input-output map can be modeled as implemeted by a network of {\it neurons}  
characterized by connectivity parameters $\vec{W}$, such that
\begin{equation}
\vec{y}=f_{\vec{W}}(\vec{x})~.
\end{equation}

\begin{figure}[htb]
\centering\includegraphics[max width=100pt]{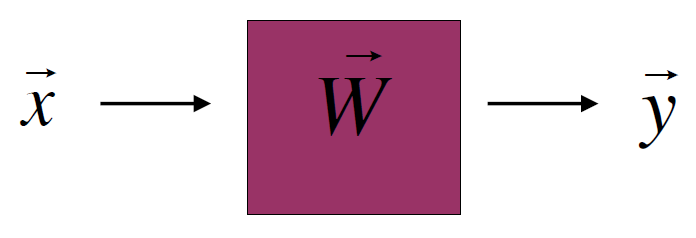}
\caption{The model as an input-output-map specified by its parameters $W$.}
\end{figure}

What specifies the value of the parameters $\vec{W}$?
The parameters of this network are determined by a learning mechanism that relies on data $\vec{\xi} \ $ that provides information about the desired input-output map:
\begin{equation}
\vec{\xi}^{\mu}=\left(\vec{x}^{\mu}, \vec{y}^{\mu}\right) \quad 1 \leq \mu \leq m~.
\end{equation}
These $m$ input-output pairs may be exact or corrupted by noise, but they are  examples of the desired map.
Supervised learning can be applied to infer a map that approximates the desired solution. 
\begin{figure}[htb]
\centering\includegraphics[max width=200pt]{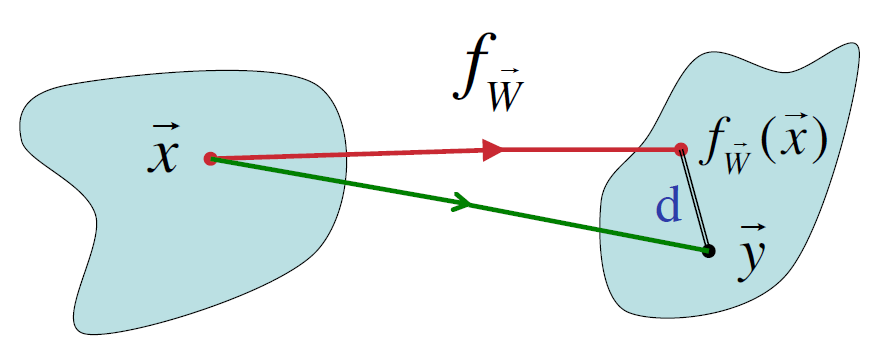}
\caption{To evaluate a model, compare  the output $f_{\vec{W}}
 (\vec{x}) \ $predicted  by the  model  to the desired output  $\vec{y}$.}
\end{figure}

The comparison between  the network's output and the desired output defines an error, and the parameters are adapted to reduce this error.
Given an example of the desired map, the error made by a specific module $\vec{W}$ on this example is
\begin{equation}
E(\vec{W} \mid \vec{x}, \vec{y})=d\left(\vec{y}, f_{\vec{W}}(\vec{x})\right)=\frac{1}{2}\left(\vec{y}-f_{\vec{W}}(\vec{x})\right)^{2}~.
\end{equation}
The loss function for such training does not need to be quadratic; it is chosen to be some kind of distance defined in output space. The choice of error function does not need to satisfy all requirements that define a proper distance: it does not need to be symmetric and it does not need to obey the triangle inequality. It only needs to satisfy that it is equal to zero if and only if its two arguments are equal. 

Given a training set of size $m$
\begin{equation}
\vec{\xi}^{\mu}=\left(\vec{x}^{\mu}, \vec{y}^{\mu}\right) \quad 1 \leq \mu \leq m~,
\end{equation}
we construct a cost function that measures the average error over the training set, the {\it learning error}:
\begin{equation}
E_{L}(\vec{W})=(1 / m) \sum_{\mu=1}^{m} E\left(\vec{W} \mid \vec{x}^{\mu}, \vec{y}^{\mu}\right)~.
\end{equation}
Most learning algorithms are based on finding the parameters $\vec{W}^{*}$ that minimize this learning error.
Most often, this is achieved by a form of gradient descent, where the weights are modified in small steps that depend in direction and size on the gradient of the learning error.
The system will then eventually converge to a local minimum of the loss landscape.
\begin{figure}[htb]
\centering\includegraphics[max width=200pt]{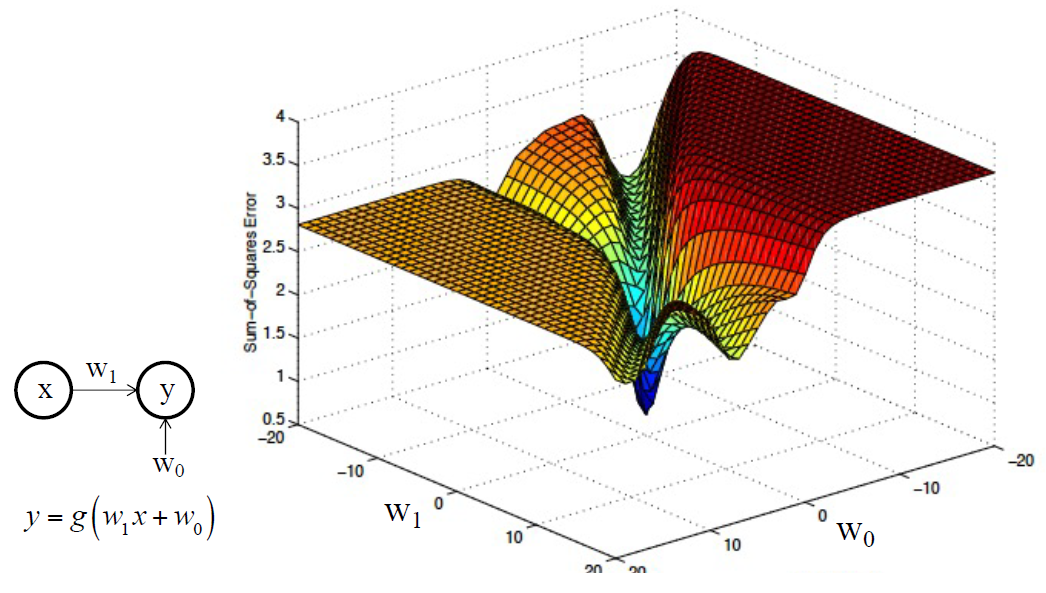}
\caption{Loss landscape with a global and local minima.}
\end{figure}

\subsection{Perceptron learning by gradient descent}
A perceptron is a network model which obtains its scalar output from an $N$-dimensional input  through a linear mapping followed by a soft nonlinearity:
\begin{equation}
y=g\left(\sum_{i=1}^{N} w_{i} x_{i}+w_{0}\right)=g\left(\vec{w}^{T} \vec{x}\right)~,
\end{equation}
where we have extended the input to $N+1$ dimensions by adding a component $x_0 =1$ to account for the bias $w_0$. The error on the $\mu$-th example and its gradient are then given by 
\begin{equation}
\quad E^{\mu}=\frac{1}{2}\left(y^{\mu}-g\left(\sum_{i=1}^{N} w_{i} x_{i}^{\mu}+w_{0}\right)\right)^{2}~,
\end{equation}
\begin{equation}
\quad \frac{\partial E^{\mu}}{\partial w_{i}}=-\left(y^{\mu}-g\left(\vec{w}^{T} \vec{x}^{\mu}\right)\right) g^{\prime}\left(\vec{w}^{T} \vec{x}^{\mu}\right) x_{i}^{\mu}~.
\end{equation}

We can then formulate the gradient descent learning as a delta rule \cite{widrow88delta}. To this end, we define the shorthand $\delta^\mu$ as in
\begin{equation}
\begin{gathered}
\frac{\partial E^{\mu}}{\partial w_{i}}=-\left(y^{\mu}-g\left(\vec{w}^{T} \vec{x}^{\mu}\right)\right) g^{\prime}\left(\vec{w}^{T} \vec{x}^{\mu}\right) x_{i}^{\mu} \equiv-\delta^{\mu} x_{i}^{\mu} \\
\delta^{\mu} \equiv g^{\prime}\left(\vec{w}^{T} \vec{x}^{\mu}\right)\left(y^{\mu}-g\left(\vec{w}^{T} \vec{x}^{\mu}\right)\right) \\
w_{i} \rightarrow w_{i}+\Delta w_{i}=w_{i}-\eta \frac{\partial E^{\mu}}{\partial w_{i}}=w_{i}+\eta \delta^{\mu} x_{i}^{\mu}~.
\end{gathered}
\end{equation}
The weight update for the perceptron is then given by
\begin{equation}
\Delta w_{i}^{\mu}=\eta \delta^{\mu} x_{i}^{\mu}~,
\end{equation}
where $\eta$ is a learning rate.

\subsection{Configuration space}

Learning in artificial neural networks can also be regarded from an ensemble viewpoint \cite{ levin90inference, solla95gibbs}. The ensemble of all possible networks compatible with a given architecture is described by its configuration space $\{\vec{W}\}$ .
This space defines all possible functions that the network can implement  given its architecture.

\begin{figure}[htb]
\centering\includegraphics[max width=100pt]{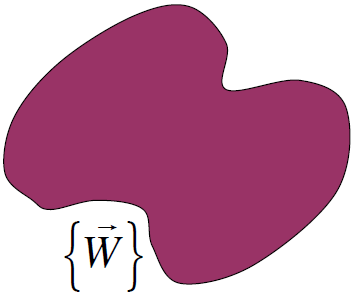}
\caption{A generic  configuration space.}
\end{figure}

The weights $\vec{W}$ follow from a  normalized prior distribution $\rho_{0}(\vec{W})$:
\begin{equation}
\int \rho_{0}(\vec{W}) d \vec{W}=1 . 
\end{equation}
This prior distribution assumes no information about the data, and should be chosen to be  as unrestricted as possible.

\subsubsection{Error-free learning}

For each example $\vec{\xi}^{\mu}=\left(\vec{x}^{\mu}, \vec{y}^{\mu}\right)$ in the training set, define a masking function $\Theta$ as
\begin{equation}
  \Theta\left(\vec{W}, \vec{\xi}^{\mu}\right)=
  \begin{cases}
    1, & \text{if}\ f_{\vec{W}}\left(\vec{x}^{\mu}\right)=\vec{y}^{\mu} \\
    0, & \text{otherwise}
  \end{cases}
\end{equation}
As we iteratively draw samples, the masking function $\Theta$ will eliminate  all network configurations that do not satisfy the constraints imposed by  the samples.
The probability distribution is modified multiplicatively:

\begin{equation}
\begin{gathered}
\rho_0(\vec{W}) \\
\Rightarrow \rho_0(\vec{W}) \ \Theta(\vec{W}, \vec{\xi}^1) \\
\Rightarrow \rho_0(\vec{W}) \ \Theta(\vec{W}, \vec{\xi}^1) \ \Theta(\vec{W}, \vec{\xi}^2)
\end{gathered}
\end{equation}

\begin{figure}[htb]
\centering\includegraphics[max width=100pt]{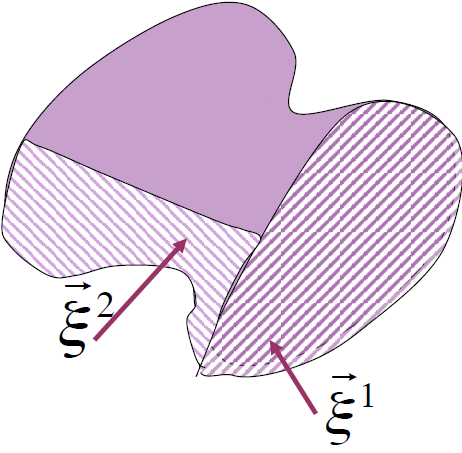}
\caption{The configuration space shrinks during learning. Configurations that do not satisfy the constraints imposed by the examples are masked away.}
\end{figure}

After the network has seen all the examples in the training set, the configuration space shrinks to
\begin{equation}
Z_m = \int d\vec{W} \rho_0(\vec{W})\prod_{\mu=1}^m \Theta(\vec{W}, \vec{\xi}^\mu)~,
\end{equation}
with 
\begin{equation}
Z_{m} \leq Z_{m-1} \leq \ldots \leq Z_{1} \leq Z_{0}=1 ,
\end{equation}
as with each iteration an additioanl subset of the configuration space may be eliminated.

\subsubsection{Learning from noisy data}
Consider an error on the $\mu$th example:
\begin{equation}
E\left(\vec{W} \mid \vec{\xi}^{\mu}\right)=d\left(\vec{y}^{\mu}, f_{\vec{W}}\left(\vec{x}^{\mu}\right)\right)
\end{equation}
If $f_{\vec{W}}\left(\vec{x}^{\mu}\right)=\vec{y}^{\mu}$, the error function vanishes and the corresponding weights are not masked away: 
$E\left(W \mid \vec{\xi}^{\mu}\right)=0 \Rightarrow \Theta\left(\vec{W}, \vec{\xi}^{\mu}\right)=1$

If $f_{\vec{W}}\left(\vec{x}^{\mu}\right) \neq \vec{y}^{\mu}$, instead of setting $\Theta\left(\vec{W}, \vec{\xi}^{\mu}\right)=0$ we can introduce a survival probability.

\begin{equation}
\Theta\left(\vec{W}, \vec{\xi}^{\mu}\right) \rightarrow \exp \left(-\beta E\left(\vec{W} \mid \vec{\xi}^{\mu}\right)\right)~.
\end{equation}
This corresponds to the difference between hard and soft masking, where configurations are attenuated by a factor exponentially controlled by the error made on the data instead of being eliminated. The exact choice of survival probability seems ad hoc at this point, but will become rigorously justified.

\begin{figure}[htb]
\centering\includegraphics[max width=100pt]{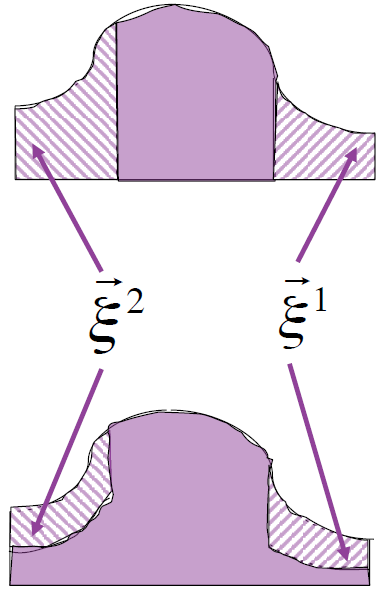}
\caption{Soft masking: Instead of eliminating subsets of configuration space, these regions  are attenuated  by an exponential factor.}
\end{figure}

The configuration space changes multiplicatively through a product of exponentials. The probability density becomes
\begin{equation}
\begin{gathered}
\rho_{0}(\vec{W}) \Rightarrow \rho_{0}(\vec{W}) \exp \left(-\beta E\left(\vec{W} \mid \vec{\xi}^{1}\right)\right)\\
\Rightarrow \rho_{0}(\vec{W}) \exp \left(-\beta E\left(\vec{W} \mid \vec{\xi}^{1}\right)\right) \exp \left(-\beta E\left(\vec{W} \mid \vec{\xi}^{2}\right)\right)\rightarrow...
\end{gathered}
\end{equation}

Given the full training set of size $m$, the effective volume of  the configuration space becomes
\begin{equation}
Z_{m}=\int d \vec{W} \rho_{0}(\vec{W}) \prod_{\mu=1}^{m} \exp \left(-\beta E\left(\vec{W} \mid \vec{\xi}^{\mu}\right)\right)
\end{equation}
Combining the product of exponentials into the exponential of the sum, we recover the mean error over all presented examples:
\begin{equation}
Z_{m}=\int d \vec{W} \rho_{0}(\vec{W}) \exp \left(-m \beta E_{L}(\vec{W})\right), 
\end{equation}
with  \begin{equation}
E_{L}(\vec{W})=(1 / m) \sum_{\mu=1}^{m}E\left(\vec{W} \mid \vec{\xi}^{\mu}\right) . 
\end{equation}
This formulation, based on the learning error and explicitly showing the size $m$ of the training set in the exponent,  will allow us to establish a helpful connection to statistical physics.

\subsubsection{Thermodynamics of learning}
In the ensemble description, the ensemble of all possible networks  is described by the prior density $\rho_{0}(W)$, and the ensemble of trained networks  is described by the posterior density $\rho_{m}(\vec{W})$ : 
\begin{equation}
\rho_{m}(\vec{W})=\frac{1}{Z_{m}} \rho_{0}(\vec{W}) \exp \left(-\beta m E_{L}(\vec{W})\right)
\end{equation}
The equation for $\rho_{m}(\vec{W})$  takes the form of a Gibbs distribution.
In comparison to  statistical physics, the noise control parameter  $\beta$ plays the role of  the inverse temperature and the learning error plays  the role of an energy function;  the latter is extensive in the sample size $m$ as opposed to the dimensionality of the configuration space.
Note that $\int d \vec{W} \rho_{m}(\vec{W})=1$, and that the partition function $Z_{m}$ provides the normalization factor. Note also that this distribution arises without invoking specific algorithms for using $E_L (\vec W)$ to explore the configuration space $\{\vec{W}\}$.

The training data $\vec{\xi}=(\vec{x}, \vec{y})$ is drawn from a natural distribution \begin{equation}
\tilde{P}(\vec{\xi})=\tilde{P}(\vec{x}, \vec{y})=\tilde{P}(\vec{y} \mid \vec{x}) \tilde{P}(\vec{x}) .
\end{equation}
Here, 
$\tilde{P}(\vec{x})$ describes the region of interest input space, and
$\tilde{P}(\vec{y} \mid \vec{x})$ describes the functional dependence.
The partition function
\begin{equation}
Z_{m}=\int d \vec{W} \rho_{0}(\vec{W}) \exp \left(-\beta \sum_{\mu=1}^{m} E\left(\vec{W} \mid \vec{\xi}^{\mu}\right)\right)
\end{equation}
depends on the specific set of data points $D=\left\{\vec{\xi}^{\mu}\right\}$ drawn from $\tilde{P}(\vec{\xi})$. 
The associated free energy
\begin{equation}
F=- \ (1 / \beta)\left\langle\left\langle\ln Z_{m}\right\rangle\right\rangle_{D}
\end{equation}
follows from averaging over all possible data sets of size $m$. The average learning error follows from the usual thermodynamic derivative:
\begin{equation}
E_{L}=- \ \frac{1}{m} \frac{\partial}{\partial \beta}\left\langle\left\langle\ln Z_{m}\right\rangle\right\rangle_{D}
\end{equation}

The entropy follows from 
\begin{equation}
F=m E_{L}-(1 / \beta) \ S
\end{equation}
For the learning process, this results in
\begin{equation}
S=-\int d \vec{W} \rho_{m}(\vec{W}) \ln \left[\frac{\rho_{m}(\vec{W})}{\rho_{0}(\vec{W})}\right]=-D_{K L}\left[\rho_{m} \mid \rho_{0}\right]~.
\end{equation}

The entropy of learning is minus the Kullback-Leibler distance between the posterior
$\rho_{m}(\bar{W})$
and the prior $\rho_{0}(W)$; this distance measures the
amount of information gained. The distance between posterior and prior increases monotonically with the size $m$ of the training set.

\begin{figure}[htb]
\centering\includegraphics[max width=200pt]{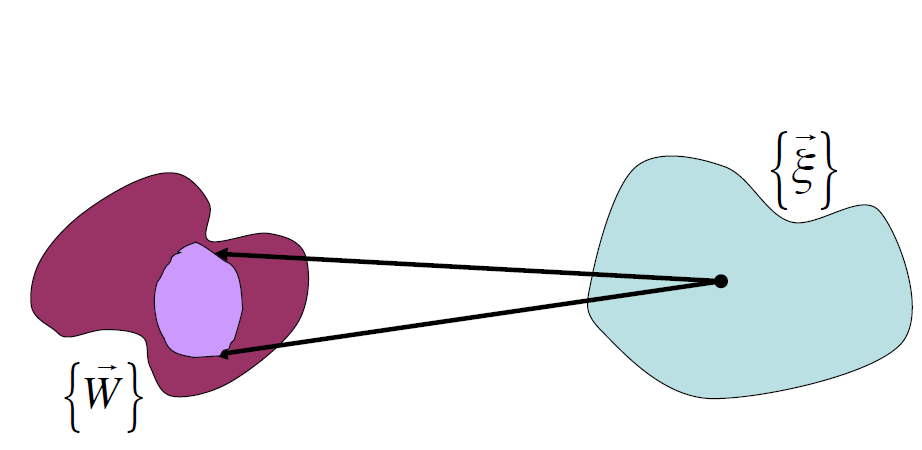}
\caption{
Relation between the configuration space of the model and the sample space. $P(\vec{W})=\rho_{0}(\vec{W})$ : prior distribution (left, dark purple). $P(\vec{W} \mid \vec{\xi})$ : distribution of hypothesis  induced by example $\vec{\xi}$ (left, light purple). 
}
\end{figure}

The entropy difference 
$\Delta H=H_{P(\vec{W})}-\left\langle\left\langle H_{P(\vec{W} \mid \bar{\vec{\xi}})}\right\rangle\right\rangle_{P(\vec{\xi})}$
equals the mutual information between the $\{\vec{W}\}$ space (the model) and the $\{\vec{\xi}\}$ space (the world) (Figs. 8 and 9). 

\subsection{Maximum likelihood learning}

There are two approaches to learning: 
minimizing the error on the data,
\begin{equation}
E_{L}(\vec{W})=\sum_{\mu=1}^{m} E\left(\vec{W} \mid \vec{\xi}^{\mu}\right)~,
\end{equation}
and maximizing the likelihood of the data,
\begin{equation}
\mathcal{L}(\vec{W})=\prod_{\mu=1}^{m} P\left(\vec{\xi}^{\mu} \mid \vec{W}\right)~.
\end{equation}

\begin{figure}[htb]
\centering\includegraphics[max width=200pt]{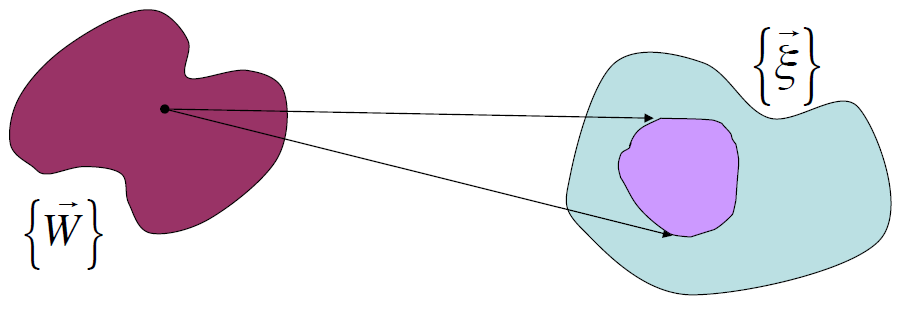}
\caption{Relation between the true sample  distribution  (right, light blue) and the distribution induced by the choice of hypothesis $\vec{W}$  (right, light purple).}
\end{figure}

The likelihood of the data given the network model can be  expressed as
\begin{equation}
\begin{gathered}
\mathcal{L}(\vec{W})=P(D \mid \vec{W})=P\left(\vec{\xi}^{1}, \vec{\xi}^{2}, \ldots, \vec{\xi}^{m} \mid \vec{W}\right)=\prod_{\mu=1}^{m} P\left(\vec{\xi}^{\mu} \mid \vec{W}\right)
\end{gathered}
\end{equation}
But, what is the form of $P(\vec{\xi} \mid \vec{W})$?
We require that these two approaches be coherent: that the minimization of the learning error implies the maximization of the data likelihood, and viceversa. This requirement leads to an explicit form for the likelihood: 
\begin{equation}
P(\vec{\xi} \mid \vec{W})=\frac{1}{z(\beta)} \exp (-\beta E(\vec{W} \mid \vec{\xi})) ,
\end{equation}
where $z(\beta)$ is a normalization factor that guarantees that $\int P(\vec{\xi} \mid \vec{W}) \ d \vec{\xi}=1$. 

We can now compute the likelihood of the data: 
\begin{align}
P(D \mid \vec{W})&=\prod_{\mu=1}^{m} P\left(\vec{\xi}^{\mu} \mid \vec{W}\right)\\
&=\frac{1}{z(\beta)^{m}} \exp \left(-\beta \sum_{\mu=1}^{m} E\left(\vec{W}  \mid \vec{\xi}^{\mu}\right)\right)\\
&=\frac{1}{z(\beta)^{m}} \exp \left(-\beta m E_{L}(\vec{W})\right)~.
\end{align}
Bayesian inference
\begin{equation}
  P(\vec{W} \mid D) = \frac{P(D \mid \vec{W}) \ P(\vec{W})}{P(D)}
\end{equation}
then leads to a  the Gibbs distribution
\begin{equation}
\rho_{m}(\vec{W})=\frac{1}{Z_{m}} \rho_{0}(\vec{W}) \exp \left(-\beta m E_{L}(\vec{W})\right)~.
\end{equation}
This is now consistent with the exponential decay of survival probability we chose before.

We can now check for the correspondence between the two formulations:

\begin{table*}[htb]
\centering
\begin{tabular}{lccc}
  & \textbf{Bayes} & $\leftrightarrow$ & \textbf{Gibbs} \\
  Prior: & $P(\vec{W})$ & $\leftrightarrow$ & $\rho_{0}(\vec{W})$ \\
  Posterior: & $P(\vec{W} \mid D)$ & $\leftrightarrow$ & $\rho_{m}(\vec{W})$ \\
  Likelihood: & $P(D \mid \vec{W})$ & $\leftrightarrow$ & $\frac{1}{z(\beta)^{m}} \exp \left(-\beta m E_{L}(\vec{W})\right)$ \\
  Evidence: & $P(D)$ & $\leftrightarrow$ & $\frac{1}{z(\beta)^{m}} Z_{m}$
\end{tabular}
\end{table*}
\noindent where $P(D)=\int d \vec{W} P(D \mid \vec{W}) P(\vec{W})$.

\subsubsection{Generalization ability}
The normalization constant $z(\beta)$ plays a role in the evaluation of prediction errors (\textit{i.e.,} has the network model  acquired a good model of the world?).

Consider now a new point $\vec{\xi}$ not part of the training data $D=\left\{{\vec{\xi}}^{1},{\vec{\xi}}^{2}, \ldots,{\vec{\xi}}^{m}\right\}$. 
What is the likelihood of this test point?
\begin{equation}
P(\vec{\xi} \mid D)=\int d \vec{W} P(\vec{\xi} \mid \vec{W}) \ P(\vec{W} \mid D)~,
\end{equation}
with
\begin{equation}
\quad P(\vec{\xi} \mid \vec{W})=\frac{1}{z(\beta)} \exp (-\beta E(\vec{W} \mid \vec{\xi}))
\end{equation}
and
\begin{equation}
P(\vec{W} \mid D)=\rho_{m}(\vec{W})=\frac{1}{Z_{m}} \rho_{0}(\vec{W}) \exp \left(-\beta \sum_{\mu=1}^{m} E\left(\vec{W} \mid \vec{\xi}^{\mu}\right)\right)~.
\end{equation}
Thus,
\begin{align}
    P(\vec{\xi} \mid D)&=\int d \vec{W} \ P(\vec{\xi} \mid \vec{W}) \  P(\vec{W} \mid D)\\
    &=\frac{1}{z(\beta) Z_{m}} \int d \vec{W} \rho_{0}(\vec{W}) \exp \left(-\beta \sum_{\mu=1}^{m+1} E\left(\vec{W} \mid \vec{\xi}^{\mu}\right)\right)~,
\end{align}
where $\vec{\xi}^{m+1}=\vec{\xi}$ is the test point; it  appears as if it had been added to the training set.
Then
\begin{equation}
P(\vec{\xi} \mid D)=\frac{Z_{m+1}}{z(\beta) Z_{m}}~.
\end{equation}

The generalization error of the trained model is defined as the logarithm of the likelihood of an arbitrary  test point $\vec{\xi}$ drawn from the same distribution as the training data:
\begin{equation}
P(\vec{\xi} \mid D)=\frac{Z_{m+1}}{z(\beta) Z_{m}} \Longrightarrow E_{G}=-\frac{1}{\beta}\left[\ln \frac{Z_{m+1}}{Z_{m}}-\ln z(\beta)\right]~.
\end{equation}
For large $m$, the difference between $\ln Z_{m+1}$ and
$\ln Z_{m}$ can be approximated by a derivative with respect to $m$. 
Then $\ln Z$ is averaged over all possible data sets of size $m$, to obtain:
\begin{equation}
E_{G}=-\frac{1}{\beta} \frac{\partial}{\partial m}\left\langle\left\langle\ln Z_{m}\right\rangle\right\rangle_{D}+\frac{1}{\beta} \ln z(\beta)~.
\end{equation}
This additional thermodynamic derivative is unusual within a canonical Gibbs ensemble. It arises because the role of the energy in the exponential factor is played by the learning error; this quantity is extensive in the number $m$ of examples in the training set, which differs from the dimensionality $D_{\vec W}$ of the configuration space. It is thus possible to take a derivative with respect to $m$ at constant $D_{\vec W}$. 

Learning and generalization errors thus correspond to the two thermodynamic derivatives:
\begin{align}
&E_{L}=- \ \frac{1}{m} \frac{\partial}{\partial \beta}\left\langle\left\langle\ln Z_{m}\right\rangle\right\rangle_{D} \\
&E_{G}=- \ \frac{1}{\beta} \frac{\partial}{\partial m}\left\langle\left\langle\ln Z_{m}\right\rangle\right\rangle_{D}+\frac{1}{\beta} \ln z(\beta)~.
\end{align}
What remains to be determined  is how to select the tolerance to errors, the inverse temperature $\beta$. In the following, we will show how this arises in a simple learning scenario.

\subsection{A simple example: The linear map}

\begin{figure}[htb]
\centering\includegraphics[max width=200pt]{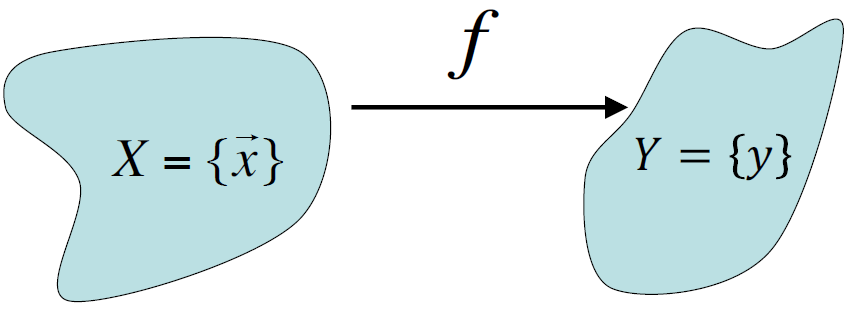}
\caption{Linear mapping $f: X \rightarrow Y$.}
\end{figure}

We consider the simplest possible network: a linear map from an $N$-dimensional input $\vec{x}$ to an scalar $y$. The parameters 
$\{\vec{W}\}$ are drawn from $\rho_{0}(\vec{W})=\mathcal{N}\left(0, C_{w}\right)$ with $C_{w}=\sigma_{w}^{2} I_{N}$ and $\sigma_{w} \gg 1$.
The map is described by
\begin{equation}
\vec{x}=\left\{x_{1}, x_{2}, \ldots, x_{N}\right\} \rightarrow y=f_{W}(\vec{x})=\sum_{i=1}^{N} W_{i} x_{i}=\vec{W}^{T} \vec{x}~.
\end{equation}

Let's assume that the inputs are also drawn from a Gaussian distribution 
$\tilde{P}(\vec{x})=\mathcal{N}\left(0, C_{x}\right)$ with $C_{x}=\sigma_{x}^{2} I_{N}$, and that  the target output for input $\vec{x}^{\mu}$ is
\begin{equation}
y^{\mu}=\vec{W}_{0}^{T} \vec{x}^{\mu}+\eta^{\mu}
\end{equation}
such that
\begin{equation}
\tilde{P}(y \mid \vec{x})=\mathcal{N}\left(\vec{W}_{0}^{T} \vec{x}^{\mu}, \sigma_{\eta}^{2}\right)~.
\end{equation}
We then train to minimize
\begin{equation}
E_{L}(\vec{W})=\frac{1}{2} \sum_{\mu=1}^{m}\left(y^{\mu}-\vec{W}^{T} \vec{x}^{\mu}\right)^{2}=\frac{1}{2} \sum_{\mu=1}^{m}\left(\left(\vec{W}-\vec{W}_{0}\right)^{T} \vec{x}^{\mu}-\eta^{\mu}\right)^{2}
\end{equation}
For $\sigma_{w} \gg 1$ and in the large $m$ limit the free energy can be computed analytically \cite {solla92levin}:
\begin{align}
\left\langle\left\langle\ln Z_{m}\right\rangle\right\rangle=&-N \ln \sigma_{w}-\frac{\mathbf{W}_{0}^{T} \cdot \mathbf{W}_{0}}{2 \sigma_{w}^{2}}-\frac{N}{2} \ln \left(2 \beta \sigma_{x}^{2} m\right) \\
&+(N-m) \beta \sigma_{\eta}^{2}+O(1 / m)~.
\end{align}
The thermodynamic derivatives are:
\begin{align}
  E_{L}&=\frac{N}{2 m \beta}+\left(1-\frac{N}{m}\right) \sigma_{\eta}^{2}+O\left(1 / m^{2}\right) \\
  E_{G}&=\frac{N}{2 m}+\beta \sigma_{\eta}^{2}+\ln \left[\left.\frac{\pi}{\beta}\right]^{1 / 2}+O\left(1 / m^{2}\right)\right.~.
\end{align}
We can now define the effective temperature $\beta_{0}$ associated to the noise in the examples as
\begin{equation}
\beta_{0}=\frac{1}{2 \sigma_{\eta}^{2}}~.
\end{equation}
The thermodynamic derivatives then simplify as
\begin{equation}
\begin{aligned}
&E_{L}=\frac{N}{2 m}\left[\frac{1}{\beta}-\frac{1}{\beta_{0}}\right]+\frac{1}{2 \beta_{0}}+O\left(1 / m^{2}\right) \\
&E_{G}=\frac{N}{2 m}+\frac{\beta}{2 \beta_{0}}+\ln \left[\frac{\pi}{\beta}\right]^{1 / 2}+O\left(1 / m^{2}\right)~.
\end{aligned}
\end{equation}

\begin{figure}[htb]
\centering\includegraphics[max width=200pt]{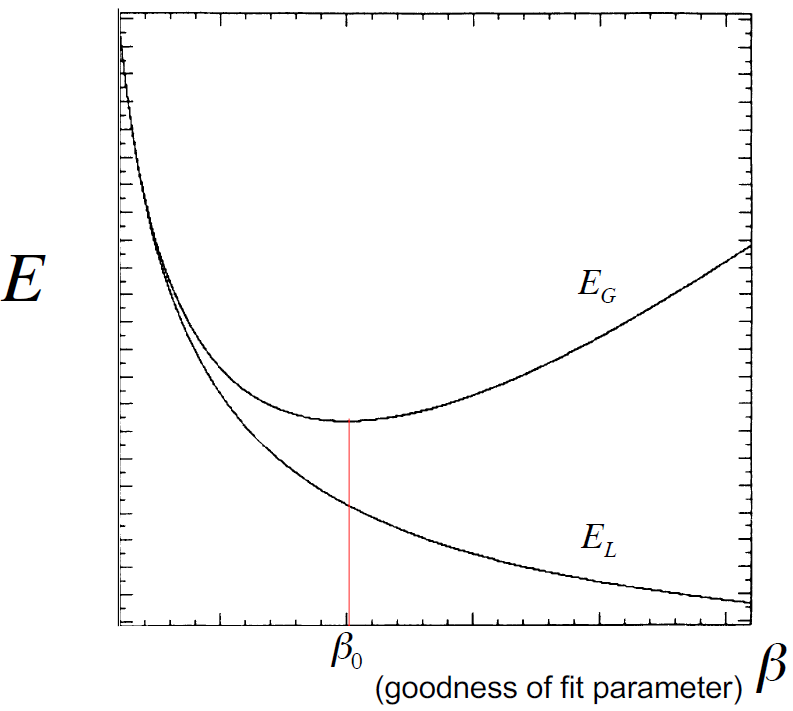}
\caption{Learning error $E_L$ and generalization error $E_G$ as function of the inverse temperature $\beta$.}
\end{figure}
For small values of $\beta$, in the high temperature regime, tolerance to learning error is large. The data is being underfit, and both $E_L$ and $E_G$ are large. As $\beta$ increases, tolerance to learning error decreases and $E_L$ decreases monotonically. While $E_G$ exceeds $E_L$ for all values of $\beta$, as expected, a different regime in $E_G$ arises for $\beta > \beta_0$. In this large $\beta$, low temperature regime, the tolerance to learning error continues to decrease but $E_G$ increases, indicating that the data is overfit. This nonmonotonic behavior in $E_G$ is observed in a variety of overfitting scenarios. Here is arises when the effective temperature of the learning algorithm falls below the value that characterizes the level of noise in the data.   
\subsection{Appendix: Derivation of the exponential form for the prediction error}
Let's require  that the minimization of the learning error
\begin{equation}
E_{L}(\vec{W})=\sum_{\mu=1}^{m} E\left(\vec{W} \mid \vec{\xi}^{\mu}\right)
\end{equation}
guarantees the maximization of the likelihood
\begin{equation}
\mathcal{L}(\vec{W})=\prod_{\mu=1}^{m} P\left(\vec{\xi}^{\mu} \mid \vec{W}\right)~.
\end{equation}
Given a training set $\left(\vec{\xi}^{1}, \vec{\xi}^{2}, \ldots, \vec{\xi}^{m}\right)$, these two functions need to be related:
\begin{equation}
\mathcal{L}(\vec{W})=\Phi\left(E_{L}(\vec{W})\right)~.
\end{equation}
Take a derivative on both sides with respect to one of the points in the training set, $\vec{\xi}^{j}$:
\begin{align}
\frac{\partial \mathcal{L}(D \mid \vec{W})}{\partial \vec{\xi}^{j}} 
&=\mathcal{L}\left(D \mid \vec{W}\right) \frac{1}{P\left(\vec{\xi}^{j} \mid \vec{W}\right)} \frac{\partial P\left(\vec{\xi}^{j} \mid \vec{W}\right)}{\partial \vec{\xi}^{j}} \\
&=\Phi^{\prime} \frac{\partial E\left(\vec{W} \mid \vec{\xi}^{j}\right)}{\partial \vec{\xi}^{j}}~.
\end{align}
This leads to
{\Large
\begin{equation}
\frac{\Phi^{\prime}}{\Phi}=\frac{\frac{1}{P\left(\vec{\xi}^{j} \mid \vec{W}\right)} \frac{\partial P\left(\vec{\xi}^{j} \mid \vec{W}\right)}{\partial \vec{\xi}^{j}}}{\frac{\partial E\left(\vec{W} \mid \vec{\xi}^{j}\right)}{\partial \vec{\xi}^{j}}}~.
\end{equation}
}
While the left-hand side of the equation depends on the full training set $\left(\vec{\xi}^{1}, \vec{\xi}^{2}, \ldots, \vec{\xi}^{m}\right)$, the right-hand side depends only on $\vec{\xi}^{j}$. The only way for this equality to hold for all values of $\left(\vec{\xi}^{1}, \vec{\xi}^{2}, \ldots, \vec{\xi}^{m}\right)$ is for both sides to be actually independent of the data, and thus equal to a constant:
{\Large
\begin{equation}
\frac{\frac{1}{P\left(\vec{\xi}^{j} \mid \vec{W}\right)} \frac{\partial P\left(\vec{\xi}^{j} \mid \vec{W}\right)}{\partial \vec{\xi}^{j}}}{\frac{\partial E\left(\vec{W} \mid \vec{\xi}^{j}\right)}{\partial \vec{\xi}^{j}}}=- \ \beta~.
\end{equation}
}
The equation
\begin{equation}
\frac{1}{P\left(\vec{\xi}^{j} \mid \vec{W}\right)} \frac{\partial P\left(\vec{\xi}^{j} \mid \vec{W}\right)}{\partial \vec{\xi}^{j}}=- \ \beta \frac{\partial E\left(\vec{W} \mid \vec{\xi}^{j}\right)}{\partial \vec{\xi}^{j}}
\end{equation}
can be integrated to obtain 
\begin{equation}
P\left(\vec{\xi}^{j} \mid \vec{W}\right) \propto \exp \left(-\beta E\left(\vec{W} \mid \vec{\xi}^{j}\right)\right)~.
\end{equation}
The normalized probability distribution is
\begin{equation}
P(\vec{\xi} \mid \vec{W})=\frac{1}{z(\beta)} \exp (-\beta E(\vec{W} \mid \vec{\xi}))~,
\end{equation}
with $z(\beta)=\int d \vec{\xi} \exp (-\beta E(\vec{W} \mid \vec{\xi}))$. Since the equation that determines $P(\vec{\xi} \mid \vec{W})$ is first order, there is only one constant of integration: $\beta$.
For $\beta>0$, minima of $E$ correspond to maxima of $P$, as required. 

\subsection{Summary}
\begin{itemize}
  \item A Gibbs probability density function describes the ensemble of trained networks; it corresponds to the Bayesian posterior. 

   \item The normalization of this probability density corresponds to the Gibbs partition function, from which a free energy can be obtained. 

   \item The thermodynamic formulation provides the learning error and a relative entropy that quantifies the exchange of information between the model and the world as a provider of examples to be learned.
   
  \item A  correspondence between error minimization and likelihood maximization provides an equation for the generalization error that leads to a novel thermodynamic derivative. 

\end{itemize}

\newpage

\section{\textsc{Lecture 2:} Generalized linear models \textit{vs.} back propagation through time}
\graphicspath{ {./Les_Houches_0722_-_Lecture_2/images/} }

We will now focus on neural networks in the biological context, and on how to analyze neural population  dynamics to find lower dimensional representations.
We start with the simultaneous recording of the activity of a population of  neurons in the brain of a nonhuman primate. High-density multi-electrode arrays (MEAs) implanted into the cortex allow for the  simultaneous recording  of the order of a hundred neurons (Fig. 12).
Given this data, we  aim to characterize the dynamics of the underlying neural circuits.
\begin{figure}[h]
\centering
\includegraphics[max width=100pt]{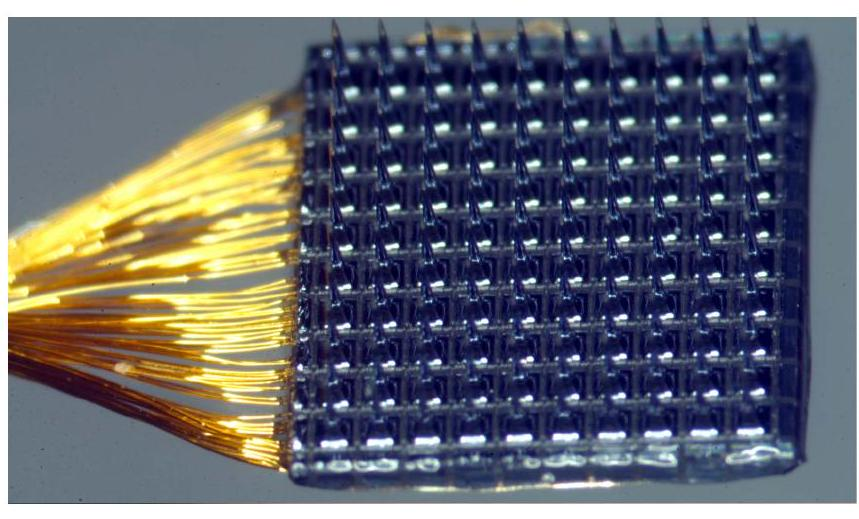}
\caption{Utah array for simutaneously  recording the activity of multiple neurons.}
\end{figure}

\subsection{Neural population activity}
Population recordings provide the temporal pattern of neuron spikes, the spike trains (Fig. 13).
To visualize these spike trains we use one row per neuron and display a dot whenever that neuron spikes. The vertical axis labels the recorded neurons, the horizontal axis is time. 
Consider a population of $N$ neurons whose spiking activity is observed during a time interval $(0, T]$.
The interval is divided into $K$ bins of size $\Delta=T / K$, labeled by an index $1 \leq k \leq K$.
In each interval $k$ we observe the number of spikes $y_{i}(k)$ emitted by neuron $i$, for all $1 \leq i \leq N$.
The number and distribution of spikes varies over neurons, over time, and depends on the settings of the experiment.
\begin{figure}[h]
  \centering
  \includegraphics[max width=200pt]{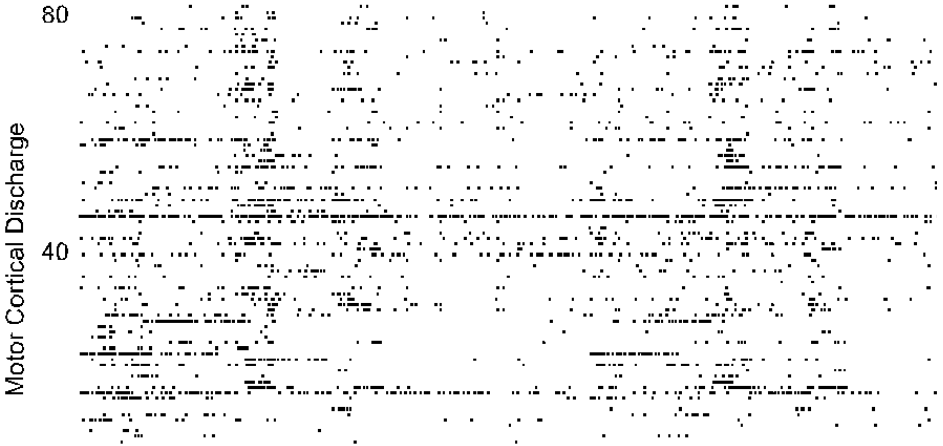}
  \caption{Neural spiking activity for a population of neurons in primary motor cortex.  }
\end{figure}

\begin{figure}[h]
\centering
\includegraphics[max width=200pt]{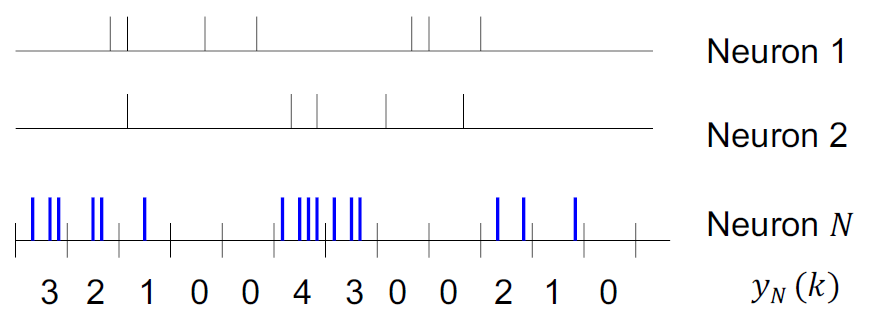}
\caption{Binned spike train.}
\end{figure}

If the subject is performing a task during the interval $(0, T]$, we search for task-specific patterns in the data.
Consider a simple motor task: \textit{center-out reaches}.
The subject is instructed to reach a visually displayed target from a center towards the outside, with an instructed  delay after the target is shown (Fig. 15).
\begin{figure}
\centering
\includegraphics[max width=300pt]{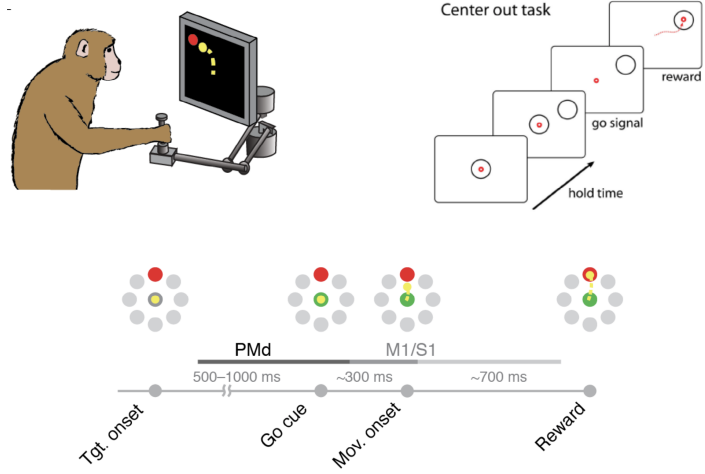}
\centering
\caption{Instructed delay center-out reaching task}
\label{fig:center_out}
\end{figure}
During the execution of this task, some  neurons in the primary motor cortex M1  exhibit a firing rate that is modulated by the direction of the reach.

\subsubsection{Analysis of neural population activity}
In our analysis we will assume that the  distribution of the spike counts $y$ within a bin can be described by a Poisson distribution:
\begin{equation}
\rho(y \mid \lambda)=\frac{\lambda^{y} e^{-\lambda}}{y !}
\end{equation}
It can be easily checked that the distribution is properly normalized
\begin{equation}
\sum_{y=0}^{\infty} \frac{\lambda^{y} e^{-\lambda}}{y !}=e^{-\lambda} \sum_{y=0}^{\infty} \frac{\lambda^{y}}{y !}=e^{-\lambda} e^{+\lambda}=1
\end{equation}
and has moments
\begin{align}
    \mathrm{E}(y)=\langle y\rangle=\lambda \\
    \operatorname{Var}(y)=\left\langle(y-\langle y\rangle)^{2}\right\rangle=\lambda
\end{align}
and Fano factor
\begin{equation}
\frac{\operatorname{Var}(y)}{\mathrm{E}(y)}=1~.
\end{equation}

\begin{figure}[h]
\centering
\includegraphics[max width=100pt]{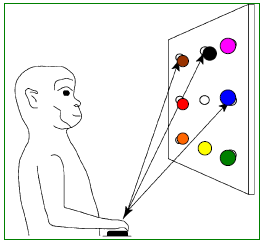}
\includegraphics[max width=150pt]{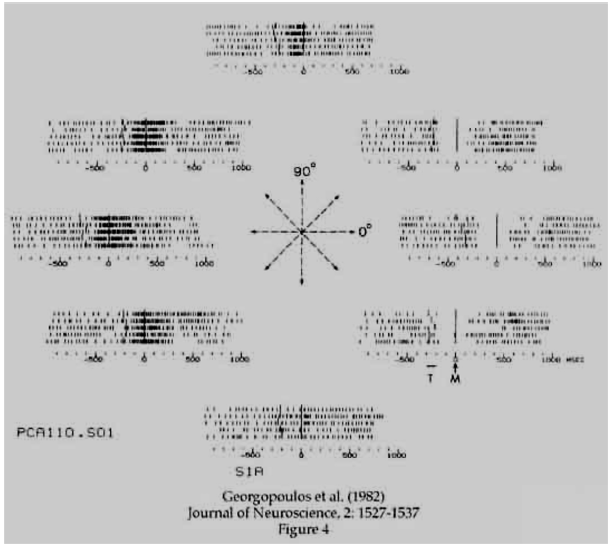}
\caption{Variability of M1 neural responses during a center-out reaching task.}
\end{figure}
Recordings from individual M1 neurons during the execution of a center-out task reveal both trial-to-trial variability and specific firing patterns that are direction dependent (Fig. 16). How can we model both variability and specificity?  For Poisson statistics, the parameter $\lambda_{i}(k)$ is the mean or expectation value of the random variable $y_{i}(k)$.
The trial-to-trial fluctuations of $y_{i}(k)$ about its mean $\lambda_{i}(k)$ describe the variability of neural activity within the $k$th time bin.
The time-dependent parameter $\lambda_{i}(k)$ provides a tool for specificity: we will model $\lambda_{i}(k)$ through its relation to sensory stimuli, motor output, and the spiking activity of other neurons.

\subsubsection{Neural activity as a Poisson process}
Let's go back to the population data $\left\{y_{i}(k)\right\}$, for $1 \leq k \leq K$ and $1 \leq i \leq N$.
The spiking activity of neuron $i$ at time interval $k$ is modeled as a Poisson process with mean $\lambda_{i}(k)$.
The probability of observing precisely $y_{i}(k)$ spikes emitted by neuron $i$ at time $k$ is given by:
\begin{equation}
P\left(y_{i}(k) \mid \lambda_{i}(k)\right)=\frac{\left(\lambda_{i}(k)\right)^{y_{i}(k)} e^{-\lambda_{i}(k)}}{y_{i}(k) !}~.
\end{equation}
What is then the probability of the observed data $\left\{y_{i}(k)\right\}$ given the parameters $\left\{\lambda_{i}(k)\right\}$?
Within the Poisson assumption, the conditional probability is given by 
\begin{equation}
P_{D}\left(\left\{y_{i}(k)\right\} \mid\left\{\lambda_{i}(k)\right\}\right)=\prod_{i=1}^{N} \prod_{k=1}^{K} \frac{\left(\lambda_{i}(k)\right)^{y_{i}(k)} e^{-\lambda_{i}(k)}}{y_{i}(k) !} ~.
\end{equation}
The log-likelihood, the logarithm of the probability, is then given by
\begin{align}
L_{D}\left(\left\{y_{i}(k)\right\} \mid\left\{\lambda_{i}(k)\right\}\right)
&=\ln \left\{\prod_{i=1}^{N} \prod_{k=1}^{K} \frac{\left(\lambda_{i}(k)\right)^{y_{i}(k)} e^{-\lambda_{i}(k)}}{y_{i}(k) !}\right\} \\
&=\left\{\sum_{i=1}^{N} \sum_{k=1}^{K}\left[y_{i}(k) \ln \lambda_{i}(k)-\lambda_{i}(k)-\ln \left(y_{i}(k) !\right)\right]\right\}~.
\end{align}
Here, $\left\{y_{i}(k)\right\}$ is the data while the neuron specific and time specific firing rates $\left\{\lambda_{i}(k)\right\}$ are the parameters of the model.
If we were to estimate the parameters $\left\{\lambda_{i}(k)\right\}$ that  maximize the likelihood of the data $\left\{y_{i}(k)\right\}$, what would we obtain? The answer is $\lambda_{i}(k) = <   y_{i}(k) >$. This answer, based on the empirical mean of the data, is not satisfactory if our goal is  to model the $\left\{\lambda_{i}(k)\right\}$.  

\subsection{A model for the time dependent and neurons specific parameters $\left\{ \lambda_{i}(k) \right\}$}

Our goal is to model the parameters $\left\{\lambda_{i}(k)\right\}$ in order to relate their values to sensory stimuli, motor outputs, and the activity of other neurons in the network.
We will approach this modeling task with generalized linear models (GLMs).
But first we will make a detour into statistics: the exponential family of probability distributions.

\subsubsection{The exponential family of probability distributions}

The exponential family encompasses probability distributions of the form:
\begin{equation}
\label{eqn:exponentialfamily}
\rho(y \mid \delta, \varphi)=\exp \left\{\frac{y \delta-b(\delta)}{a(\varphi)}+c(y, \varphi)\right\}~.
\end{equation}
Here, $y$ is the random variable whose probability density function is given by $\rho$.
The distribution is parametrized by the canonical parameter $\delta$ and the dispersion parameter $\varphi$.
The functions $a$, $b$, and $c$ need to be specified; they define the various distributions within the family.

The term $c(y, \varphi)$ plays an important role: it provides a normalization function that guarantees $\int d y \ \rho(y \mid \delta, \varphi)=1$ for all $\delta, \varphi$.
Since $\int d y  \ \rho_{y}(y \mid \delta, \varphi)=1$ for all $\delta, \varphi$ then:
\begin{equation}
\begin{array}{lll}
\frac{\partial}{\partial \delta} \int d y \rho_{y}(y \mid \delta, \varphi)=0 & \Rightarrow & \mathrm{E}(y)=b^{\prime}(\delta) \\
\frac{\partial^{2}}{\partial \delta^{2}} \int d y \rho_{y}(y \mid \delta, \varphi)=0 & \Rightarrow & \operatorname{Var}(y)=a(\varphi) b^{\prime \prime}(\delta)~.
\end{array}
\end{equation}
Note that the canonical parameter $\delta$ fully determines the mean $\mathrm{E}(y)$ through $b(\delta)$, while the variance $\operatorname{Var}(y)$ requires additional information provided by the dispersion parameter through $a(\varphi)$ \cite{mccullagh2019generalized}.

Consider the family of canonical exponential distributions with canonical parameter $\delta$ and dispersion parameter $\varphi$. Why is this family important? 
Because the normal, Bernoulli, binomial, multinomial, Poisson, gamma, geometric, chi-square, beta, and a few other distributions are all members of this exponential family.

\subsubsection{The Poisson distribution as a member of the exponential family}

Let's have a closer look at one specific example, the Poisson distribution:
\begin{equation}
\rho(y \mid \lambda)=\frac{\lambda^{y} e^{-\lambda}}{y !}=\exp \{y \ln \lambda-\lambda-\ln (y !)\}~.
\end{equation}
This takes the form of the exponential family (Eq. \ref{eqn:exponentialfamily}) for $a(\varphi)=1, \delta=\ln \lambda, b(\delta)=\lambda$, and $c(\varphi, y)=-\ln (y !)$.

The relations $\mathrm{E}(y)=b^{\prime}(\delta)$ and $\operatorname{Var}(y)=a(\varphi) b^{\prime \prime}(\delta)$
hold for any probability density function within the exponential family. 
When applied to the Poisson case, they imply:
\begin{equation}
\begin{aligned}
  \mathrm{E}(y)&=b^{\prime}(\delta)=\lambda \\
  \operatorname{Var}(y)&=a(\varphi) b^{\prime \prime}(\delta)=\lambda~.
\end{aligned}
\end{equation}
For Poisson statistics, $\mathrm{E}(y)=\operatorname{Var}(y)=\lambda$ implies:
\begin{equation}
b^{\prime}(\delta)=a(\varphi) b^{\prime \prime}(\delta) \implies \left\{\begin{array}{l}
a(\varphi)=1 \\
b(\delta)=e^{\delta}=\lambda
\end{array}\right.~.
\end{equation}

In a generalized linear model (GLM) for a probability distribution that is a member of the exponential family \cite{mccullagh2019generalized}, the expectation value $\mathrm{E}(y)$ is related to the canonical parameter $\delta$ via a nonlinear link function $g$:
\begin{equation}
g(\mathrm{E}(y))=\delta \quad \quad \mathrm{E}(y)=g^{-1}(\delta)~.
\end{equation}
This is the only nonlinearity in the model, as the canonical parameter $\delta$ is constructed as a linear combination of all observed variables that can \textit{explain} the random variable $y$.
In the Poisson case, $\delta=\ln \lambda=\ln  (\mathrm{E}(y))$, and the nonlinear link function $g$ is the logarithm:
\begin{equation}
\begin{gathered}
\lambda=\mathrm{E}(y)=g^{-1}(\delta)=\exp (\delta) \\
\delta=g(\lambda)=\ln (\lambda)
\end{gathered}
\end{equation}

\subsection{Generalized linear models for spiking neurons}

The parameter $\lambda_{i}(k)$ is the time-dependent and neuron-specific mean of a Poisson process.
In a GLM for a Poisson distribution, it is the logarithm (link function) of the mean that is expressed as a linear combination of all observed variables that can be used to explain the observed firing rates.
Here, we have
\begin{itemize}
    \item Internal covariates: preceding neural activity (hidden neurons)
    \item External covariates: sensory stimulus (input), direction of motion (output)
\end{itemize}

We know the spiking history of the ensemble of $N$ neurons up to the current time $t$. We denote this as the spiking history of the ensemble:
\begin{equation}
H(t)=\left\{\left\{y_{i}\left(t^{\prime}\right)\right\}_{i=1}^{N}, t^{\prime} \leq t\right\}
\end{equation}
Given this information, what is our expectation of the number of spikes that neuron $i$ will fire in the interval $(t, t+\Delta)$? 
This is the \emph{conditional intensity} $\lambda_{i}(t \mid H(t))$, a strictly positive function that provides a history-dependent generalization of the time dependent rate of an inhomogeneous Poisson process \cite{truccolo2005eden}.

\subsubsection{A GLM based on internal covariates}

 The history-dependent model for $\lambda_{i}(t \mid H(t))$ in the generalized linear model (GLM) approach is given by:

\begin{equation}
\delta_{i}(t \mid H(t),\{\alpha\})=\ln \lambda_{i}(t \mid H(t),\{\alpha\}) = 
\alpha_{i 0}+\sum_{j=1}^{N} \sum_{m=1}^{\tau_{N}} \alpha_{i j}(m) y_{j}(t-m)~, 
\end{equation}
which leads to a linear-nonlinear model for $\lambda_{i}(t)$:
\begin{equation}
\lambda_{i}(t \mid H(t),\{\alpha\})=\exp \left\{\alpha_{i 0}+\sum_{j=1}^{N} \sum_{m=1}^{\tau_{N}} \alpha_{i j}(m) y_{j}(t-m)\right\}~.
\end{equation}
Once the model model parameters $\left\{ \alpha_{i j}(m) \right\}$ have been specified, this equation for $\lambda_i(t)$ provides a generative model (Fig. 17). The kernel parameter $\alpha_{i j}(m)$ quantifies the effect that the spiking activity of neuron $j$ at time bin $(t-m)$ has on the spiking activity of neuron $i$ at time bin $t$. For a given $(i,j)$ pair, the time dependent $\alpha_{i j}(m)$ defines an effective connectivity (Fig. 18).  

\begin{figure}
\centering
\includegraphics[max width=150pt]{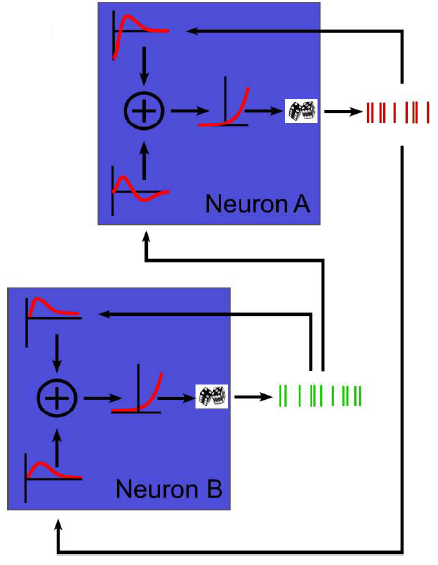}
\caption{History dependent GLM model for the time-dependent neuron-specific mean firing rate.}
\end{figure}
\begin{figure}
\centering
\includegraphics[max width=150pt]{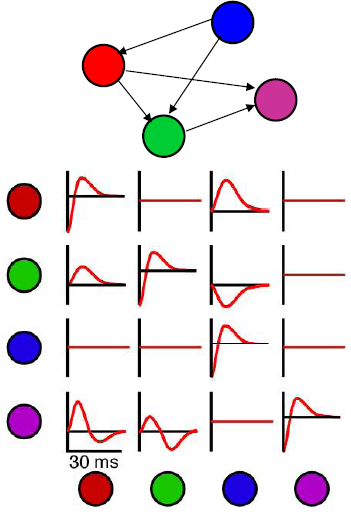}
\caption{Effective connectivity.}
\end{figure}

How are the model parameters $\{\alpha_{i j}(m) \}$ determined? Given the data $\left\{y_{i}(k)\right\}$, we search for parameters that maximize the likelihood of the data:
\begin{equation}
L_{D}\left(\left\{y_{i}(k)\right\}\right) \propto\left\{\sum_{i=1}^{N} \sum_{k=1}^{K}\left(y_{i}(k) \ln \lambda_{i}(k)-\lambda_{i}(k)\right)\right\}~.
\end{equation}
A term that does not depend on $\left\{\lambda_{i}(k)\right\}$ and thus does not depend on the parameters has been dropped. Explicitly, the likelihood function to be maximized is:
\begin{equation}
\begin{aligned}
L_{D}\left(\left\{y_{i}(k)\right\} \mid \{\alpha\}\right) 
  \propto & \left\{\sum_{i=1}^{N} \sum_{k=1}^{K}\left(y_{i}(k)\left[\alpha_{i 0}+\sum_{j=1}^{N} \sum_{m=1}^{\tau_{v}} \alpha_{i j}(m) y_{j}(k-m)\right] \right.\right. \\
  &\left.\left.-\exp \left[\alpha_{i 0}+\sum_{j=1}^{N} \sum_{m=1}^{\tau_{N}} \alpha_{i j}(m) y_{j}(k-m)\right]\right)\right\}~.
\end{aligned}
\end{equation}

\subsubsection{Learning a history dependent GLM by gradient ascent}

The maximization of the likelihood proceeds via gradient ascent with an adaptive step size. 
To implement this algorithm, we need to compute the first and second derivatives of the likelihood with respect to the parameters $\{\alpha_{ij} (m)\}$.
The gradient that drives the uphill search is given by:
\begin{equation}
\left.\frac{\partial L_{D}\left(\left\{y_{i}(k)\right\} \mid\{\alpha\}\right)}{\partial \alpha_{i j}(m)}\right|_{(\mu)}=\sum_{k=1}^{K}\left\{\left[y_{i}(k)-\langle y_{i}(k) \rangle^{(\mu)}\right] y_{j}(k-m)\right\}
\end{equation}
The update of the parameter $\alpha_{i j}(m)$ is given by the product of the activity $y_{j}(k-m)$ of the {\it presynaptic} neuron $j$ at time lag $m$ and the difference between the actual activity $y_{i}(k)$ of the {\it postsynaptic} neuron and our current estimate of it at iteration $(\mu)$. The rule {\it presynaptic activity times postsynaptic error} is a famous learning rule, called the Delta Rule \cite{widrow88delta}. We italicize {\it presynaptic} and {\it postsynaptic} to avoid the implication that the parameter $\alpha_{i j}(m)$ is an actual synaptic strength.

The components of the Hessian matrix of second derivatives that controls the size of the uphill steps are given by:
\begin{equation}
\left.\frac{\partial^{2} L_{D}\left( \left\{y_{i}(k)\right\} \mid \{\alpha\}\right)}{\partial \alpha_{i j}(m) \ \partial \alpha_{i j'}\left(m^{\prime}\right)}\right|_{(\omega)}=-\sum_{k=1}^{K} \langle y_{i}(k) \rangle^{(\mu)} y_{j}(k-m) y_{j^{\prime}}(k-m^{\prime}).
\end{equation}
Now there are two {\it presynaptic} neurons: neuron $j$ at time lag $m$ and neuron $j^{\prime}$ at time lag $m^{\prime}$. 
Their activities are multiplied, and this product is weighted by our current estimate of the activity of the {\it postsynaptic} neuron $i$.
Note the overall minus sign! The variables $\{y\}$ represent number of spikes emitted during a bin of size $\Delta$. These variables and their averages are always non-negative. Every component of the Hessian matrix is negative - the surface is everywhere convex!

The gradient ascent algorithm can be written as follows:
\begin{equation}
\vec{\alpha}^{(\mu+1)}=\vec{\alpha}^{(\mu)}+\mathrm{E}^{(\mu)} \vec{\nabla} L^{(\mu)}
\end{equation}
Here, $\vec{\alpha}$ is a listing of all the parameters needed to specify the model; $\vec{\nabla} L$ is the gradient of the likelihood function $L$, obtained by taking the derivative of $L$ with respect to each  parameter in $\vec{\alpha}$; and $\mathrm{E}$ is the matrix of step sizes, obtained by inverting the Hessian matrix of second derivatives of the likelihood function. If the model requires $p$ parameters, then both $\vec{\alpha}$ and $\vec{\nabla} L$ are $p$-dimensional vectors, and $\mathrm{E}$ is a $p \times p$ matrix.

\subsection{An example: a GLM for the center-out task}

\begin{figure*}
\centering
\includegraphics[max width=179pt]{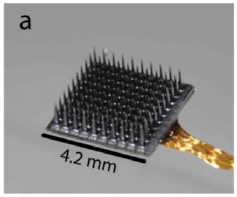}
\includegraphics[max width=160pt]{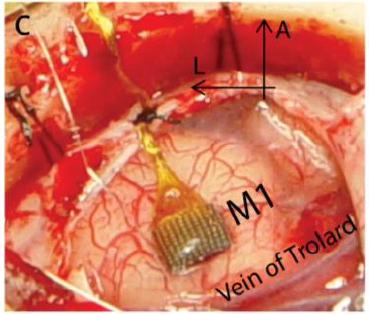}
\centering
\caption{A multi-electrode array (left) that allows  for the simultaneous recording of about a hundred neurons is shown when inserted in primary motor cortex M1 (right).}
\end{figure*}

We show the analysis of data acquired from two humans with tetraplegia while they participated in a  clinical trial. The data consists of M1 recordings (Fig. 19) obtained while the participants performed the center-out task; the (x,y) motion of the cursor followed from decoding M1 neural activity \cite{truccolo2010collective}.

 \subsubsection{A GLM based on spiking history for the center-out task}
 
To formulate the GLM, we ask what is the probability that neuron $i$ spikes at bin $k$, conditioned on the spiking history $H_{k}$ of the neural population during the preceding $100 \mathrm{~ms}$:
\begin{equation}
 \ln \lambda_{i}\left(k \mid H_{k}\right)=\alpha_{i 0}+\sum_{m=1}^{\tau_{N}} \alpha_{i i}(m) y_{i}(k-m)+\sum_{j=1, j \neq i}^{N} \sum_{m=1}^{\tau_{N}} \alpha_{i j}(m) y_{j}(k-m)~.
\end{equation}
Here, $\Delta=1 \mathrm{~ms}$ and $\tau_{N}=100$. Data is used to fit the conditional intensity $\lambda_{i}(k)$ and obtain the background level $\alpha_{i 0}$ of spiking activity for neuron $i$, the kernel $\alpha_{i i}(m)$ related to its intrinsic history effects, and the kernels $\alpha_{i j}(m)$ related to history effects due to the other neurons $j \neq i$.
Once the model is fitted, the estimated probability of a spike at any time bin can be computed.

The fitting of the model parameters involves the following steps:

\noindent (1) Given the data $\left\{y_{i}(m)\right\}$ and the current value $\left\{\alpha^{(\mu)}\right\}$ of the parameters, construct:
\begin{equation}
 \langle y_{i}(k) \rangle^{(\mu)}=\exp \left[\alpha_{i 0}^{(\mu)}+\sum_{j^{\prime}=1}^{N} \sum_{m^{\prime}=1}^{\tau_{N}} \alpha_{i j^{\prime}}^{(\mu)}(m^{\prime}) y_{j^{\prime}}(k-m^{\prime})\right]
\end{equation}
Once the estimates $\langle y_{i}(k) \rangle^{(\mu)}$ have been computed, the current values of the parameters are no longer needed. 

\noindent (2) Build the components of the gradient vector:
\begin{equation}
\left.\frac{\partial L_{D}\left(\left\{y_{i}(k)\right\} \mid\{\alpha\}\right)}{\partial \alpha_{i j}(m)}\right|_{(\mu)}=\sum_{k=1}^{K}\left[y_{i}(k)- \langle y_{i}(k) \rangle^{(\mu)}\right] y_{j}(k-m)
\end{equation}
(3) Build the components of the Hessian matrix:
\begin{equation}
 \left.\frac{\partial^{2} L_{D}\left(\left\{y_{i}(k)\right\} \mid\{\alpha\}\right)}{\partial \alpha_{i j}(m) \partial \alpha_{i j^{\prime}}\left(m^{\prime}\right)}\right|_{(\mu)}=-\sum_{k=1}^{K} \left\langle y_{i}(k) \rangle^{(\mu)} y_{j}(k-m) y_{j^{\prime}}\left(k-m^{\prime}\right)\right.
\end{equation}
(4) Invert the Hessian matrix of second derivatives to obtain the matrix Epsilon of step sizes:
\begin{equation}
\mathrm{E}=-H^{-1}
\end{equation}
(5) Multiply the matrix $\mathrm{E}$ and the gradient $\vec{\nabla} L$ to obtain the update:
\begin{equation}
\vec{\alpha}^{(\mu+1)}=\vec{\alpha}^{(\mu)}+\mathrm{E}^{(\mu)}  \ \vec{\nabla} L^{(\mu)}
\end{equation}
Examples of fitted temporal filters are shown for $i=34$ using $\gamma_{i j}(m)=\exp \left(\alpha_{i j}(m)\right)$ (Fig. 20).
\begin{figure*}
\centering
\includegraphics[max width=150pt]{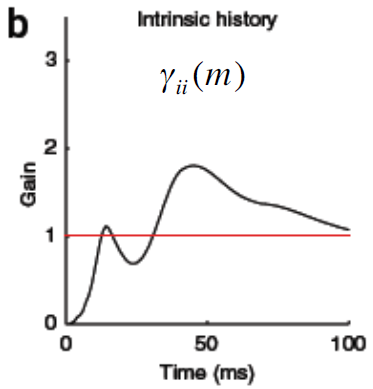}
\includegraphics[max width=150pt]{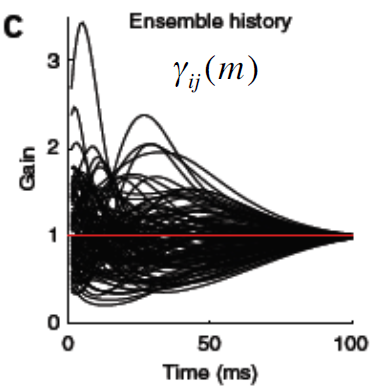}
\centering
\caption{$i=34$ \hspace{3.25cm} $i=34, j \neq i$ \hspace{1cm}}
\label{fig:gen-model}
\end{figure*}

\subsubsection{A GLM based on direction of motion for the center out task}

We now wish to use information about the direction of motion in order to predict the mean $\lambda_i(t)$ of the Poisson process.  Information about a planar reach of extent $r$ in  a direction characterized by $\theta$ can be extracted from the firing activity of orientation  selective M1 neurons \cite{georgopoulos1982relations}: 
\begin{equation}
f_{i} (r, \theta) =b_{i}+r a_{i}(1 / 2)\left[1+\cos \left(\theta-\theta_{i}\right)\right]~,
\end{equation}
where $b_{i}$ is the background activity, $a_{i}$ is the amplitude of activity modulation, and $\theta_{i}$ is the preferred direction of neuron $i$ (Fig. 21).

\begin{figure*}[h]
\centering
\includegraphics[max width=200pt]{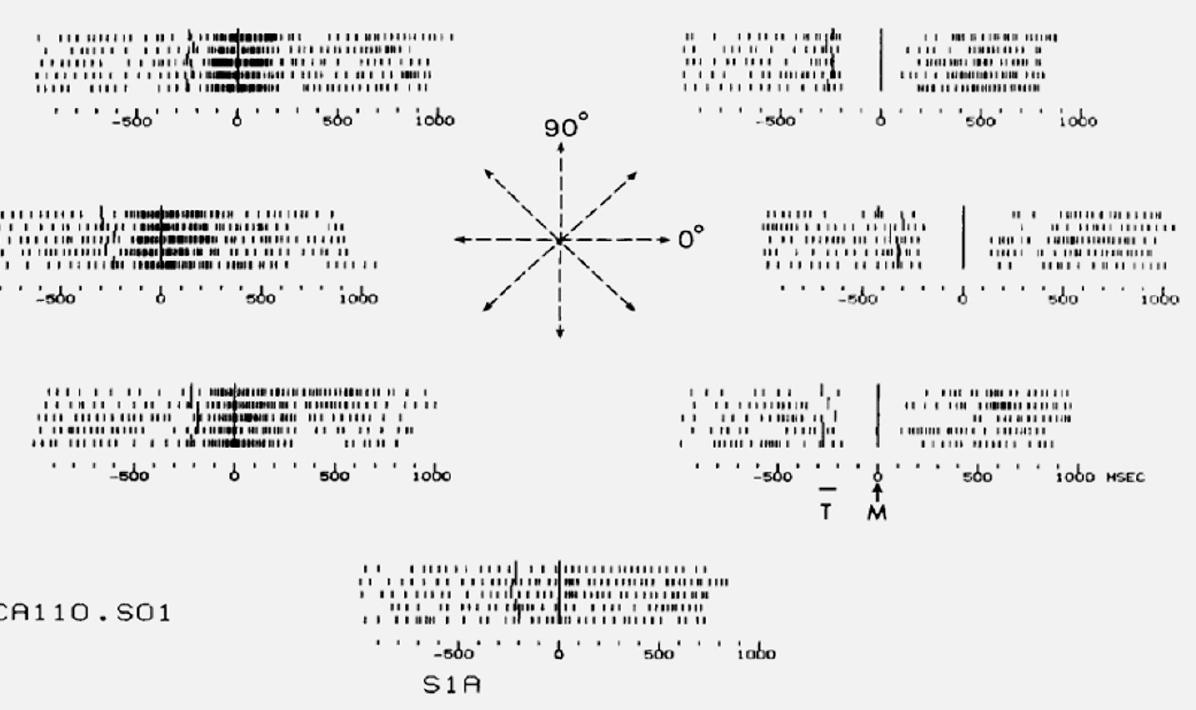}
\includegraphics[max width=165pt]{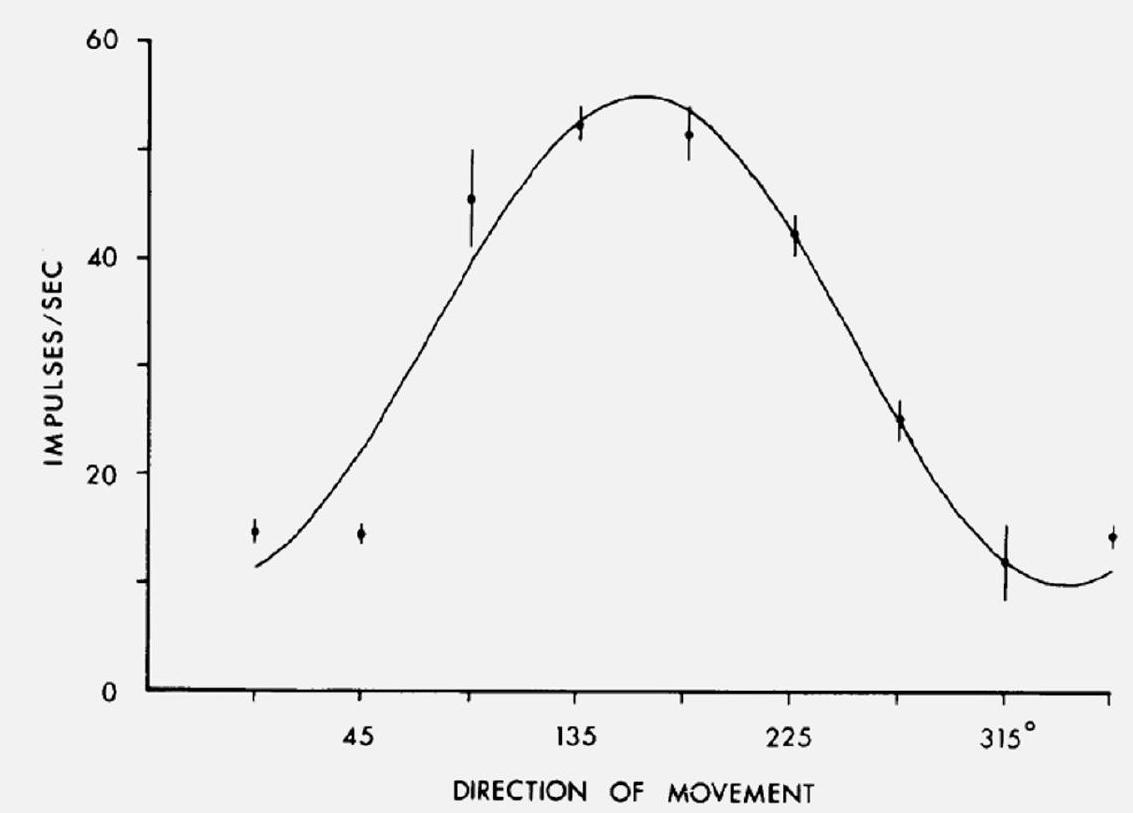}
\centering
\caption{The tuning curve is characterized by three model parameters. For experimental details, see \cite{georgopoulos1982relations}.}
\end{figure*}

This observation would allow to construct a hierarchy of three increasingly detailed GLM models:

\noindent(1) Non-interacting cosine-tuned neurons:
\begin{equation}
\ln \lambda_{i}(t)=\alpha_{i 0}+(1 / 2) \alpha_{i R} r\left(t+\tau_{R}\right)\left[1+\cos \left(\theta\left(t+\tau_{R}\right)-\theta_{i}\right)\right]
\end{equation}
(2) History-dependent non-interacting cosine-tuned neurons:
\begin{equation}
\ln \lambda_{i}(t)=\alpha_{i 0}+(1 / 2) \alpha_{i R} r\left(t+\tau_{R}\right)\left[1+\cos \left(\theta\left(t+\tau_{R}\right)-\theta_{i}\right)\right]
+\sum_{m=1}^{\tau_{N}} \alpha_{i i}(m) y_{i}(t-m)
\end{equation}
(3) History-dependent interacting cosine-tuned neurons:
\begin{equation}
\ln \lambda_{i}(t)=\alpha_{i 0}+(1 / 2) \alpha_{i R} r\left(t+\tau_{R}\right)\left[1+\cos \left(\theta\left(t+\tau_{R}\right)-\theta_{i}\right)\right]+
\sum_{j=1}^{N} \sum_{m=1}^{\tau_{N}} \alpha_{i j}(m) y_{j}(t-m)
\end{equation}

\subsection{GLM models of increasing complexity}

As an illustration of this approach, we review the analysis of MEA (MultiElectrode Array) recordings of neural activity in the arm area of primary motor cortex (M1) of awake and behaving monkeys involved in the execution of a two-dimensional tracking task: the pursuit and capture of a smoothly and randomly moving visual target \cite{truccolo2005eden}. The target was tracked by moving a two-link low friction manipulandum that constrained hand movementx to the horizontal plane. The $(\mathrm{x}, \mathrm{y})$ hand position was digitized and resampled at $1 \mathrm{~KHz}$; low-pass filtered finite differences of position were used to obtain the two components of velocity. 

\subsubsection{Two GLM models for the tracking task}

Two models were used to analyze the data:
\\

\noindent (1) Velocity model \cite{paninski2004spatiotemporal} 
\\

\noindent (2) Velocity model plus intrinsic spiking history \cite{truccolo2005eden} \\

The velocity model uses $v_{x}(t+\tau_{R})$ and $v_{y}(t+\tau_{R}$) as explanatory variables for $\lambda_{i} (t)$:
\begin{equation}
\begin{split}
& \lambda_{i}\left(t \mid v\left(t+\tau_{R}\right), \theta\left(t+\tau_{R}\right),\left\{\alpha_{i 0}, \alpha_{i X}, \alpha_{i Y}\right\}\right) \\
& = \exp \left\{\alpha_{i 0}+\alpha_{i X} v_{X}\left(t+\tau_{R}\right)+\alpha_{i Y} v_{Y}\left(t+\tau_{R}\right)\right\} \\
& = \exp \left\{\alpha_{i 0}+v\left(t+\tau_{R}\right)\left[\alpha_{i X} \cos \left(\theta\left(t+\tau_{R}\right)\right)+\alpha_{i Y} \sin \left(\theta\left(t+\tau_{R}\right)\right)\right]\right\}
\end{split} 
\end{equation}

Note that the model does not include a sum over time lags; it uses single time shift $\tau_{R}=150 \mathrm{~ms}$. There are only three parameters to be determined through a maximum likelihood fit to the spiking data of the $i$th neuron: $\left\{\alpha_{i 0}, \alpha_{i X}, \alpha_{i Y}\right\}$.
Once the values for these parameters have been specified, the conditional intensity $\lambda_{i}(t)$ can be plotted as a function of the subsequent velocity in polar coordinates: $v\left(t+\tau_{R}\right), \theta\left(t+\tau_{R}\right)$ (Fig. 22).

\begin{figure*}
\centering
\includegraphics[max width=350pt]{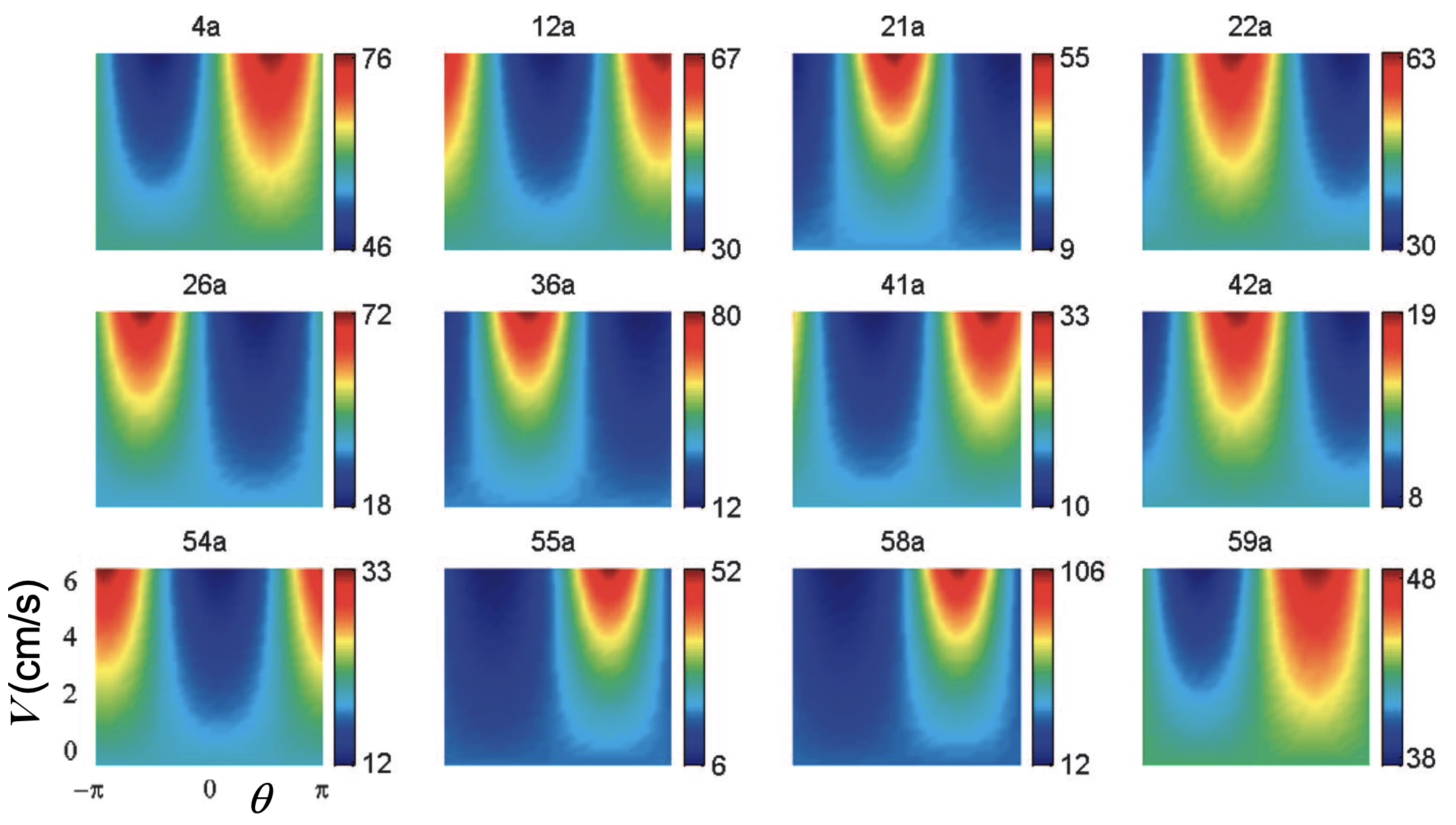}
\centering
\caption{Velocity tuning functions for 12 different neurons. The values of the mean number of spikes $\lambda$ are color coded.}
\end{figure*}

The next level of model complexity corresponds to including the intrinsic spiking history as an internal covariate:
\begin{equation}
\begin{split}
& \lambda_{i}\left(t \mid H_{i}(t), v\left(t+\tau_{R}\right), \theta\left(t+\tau_{R}\right),\left\{\alpha_{i 0},\left\{\alpha_{i i}(m)\right\}, \alpha_{i X}, \alpha_{i Y}\right\}\right)\\
& = \exp \left\{\alpha_{i 0}+v\left(t+\tau_{R}\right)\left[\alpha_{i X} \cos \left(\theta\left(t+\tau_{R}\right)\right)+\alpha_{i Y} \sin \left(\theta\left(t+\tau_{R}\right)\right)\right]+\sum_{m=1}^{\tau_{N}} \alpha_{i i}(m) y_{i}(t-m)\right\}~.
\end{split}
\end{equation}
In addition to the three parameters $\left\{\alpha_{i 0}, \alpha_{i x}, \alpha_{i Y}\right\}$, this model for $\lambda_{i}(t)$ includes the parameters $\left\{\alpha_{i i}(m)\right\}$ that characterize the intrinsic history filter.

This model was implemented by using a bin size  $\Delta=1 \mathrm{~ms}$. At this time resolution, the number of spikes $y_{i}(t-m)$ in any bin can only be $0$ or $1$. The maximum temporal length of the intrinsic history filter was set to $\tau_{N}=120$. Results for neuron $i=$75a (Fig. 23) show distinctive velocity tuning. The intrinsic history filter shows that history effects extended only 60 ms into the past. The refractory period, corresponding to negative filter coefficients,   lasts about 18 ms after a spike. The firing probability then increases and peaks at about 30 ms after a spike. 


\begin{figure*}
\centering
\includegraphics[max width=200pt,scale=0.9]{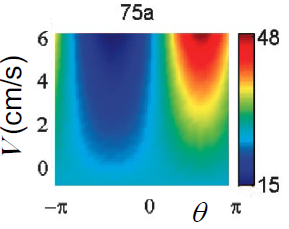}
\includegraphics[max width=150pt]{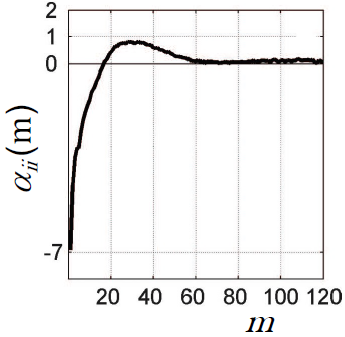}
\centering
\caption{Velocity tuning function (left) and intrinsic history filter (right) for neuron $75 \mathrm{a}$.}
\end{figure*}

\subsubsection{Time rescaling for model comparison}

To which extent does the inclusion of an intrinsic history filter result in model improvement?

\begin{figure}[h]
\centering
\includegraphics[max width=150pt,scale=0.9]{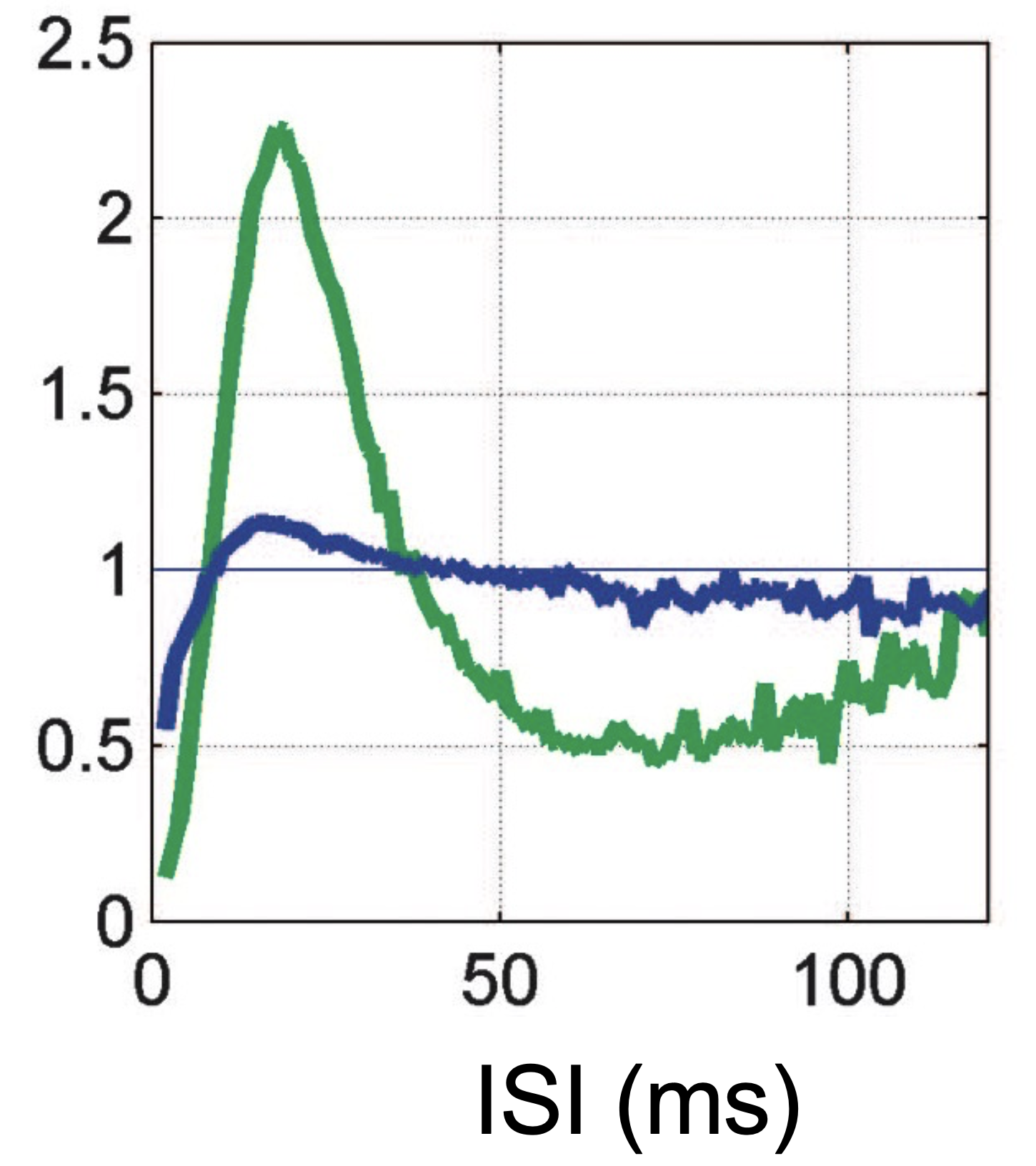}
\centering
\caption{Mean ratio of observed to
expected $z_j$ values indicates that the velocity only model (green) overestimates $\lambda_{i} (t)$
for about 10 ms after a spike and underestimates $\lambda_{i} (t)$ for longer times.}
\end{figure}

For nested models such as considered here, the {\it time-rescaling theorem} \cite{brown2002timerescaling, truccolo2005eden}
provides a useful tool for addressing this question. Given the spike times $\{ t_{j} \}$ of all spikes emitted by a specific neuron  during the $(0,T]$ observational period, we compute for each interspike interval $\tau_{j}= t_{j+1}-t_{j}$ the quantity $z_j$ defined as
\begin{equation}
    z_j = 1 - \exp \left(-\int_{t_{j}}^{t_{j+1}} \lambda (t) \ dt\right)~.
\end{equation}
For a Poisson process, the interspike interval $\tau_{j}$ is a random variable with an exponential probability density in the $(0, \infty)$ interval. If $\lambda(t)$ is the correct mean firing rate for this process, $z_j$ will be a random variable with uniform probability density in the $(0,1)$ interval. 

To implement the time-rescaling test, we compute the empirical probability density of the $z_j$ values that result from the observed spike times and the fitted model for $\lambda(t)$, and calculate the ratio of the empirical to the theoretical uniform density. When plotted as a function of the corresponding values of $\tau_j$ (Fig. 24), this ratio clearly demonstrates the modeling improvement achieved by incorporating the intrinsic spike histoty (blue curve). The velocity model (green curve) overestimates $\lambda_{i} (t)$ for 10 ms after a spike, and it subsequently  
underestimates  $\lambda_{i} (t)$. The negative (positive) coefficients of the intrinsic history filter almost completely eliminate the over (under) estimation of the $\lambda_{i} (t)$ based on
the velocity model alone.

\subsection{Summary}
\begin{itemize}
  \item Generalized linear models provide a principled and systematic approach to modeling the time-dependent rate of inhomogenous Poisson processes that describe the expected firing activity of a neural population.

  \item The logarithm of the time-dependent rate for each neuron is modeled as a linear combination of intrinsic (the preceding firing activity of all neurons in the population) and extrinsic (the preceding input stimulus or subsequent output activity) observables.

  \item The likelihood of the data is log-convex; optimal model parameters follow from an unambiguous gradient ascent algorithm.

\end{itemize}

\newpage

\section{\textsc{Lecture 3:} Linear and nonlinear dimensionality reduction}
\graphicspath{ {./Les_Houches_0722_-_Lecture_3/images/} }

Let's go back to a motor task that we have already used as an example: the center-out reaching task (Fig. 25). Each trial involves a reach from the center position to one of eight targets on a circle. After target presentation, the subject waits for the go cue to execute to reach. Each trial takes a few seconds. 
\begin{figure}[h]
\centering
\includegraphics[max width=300pt]{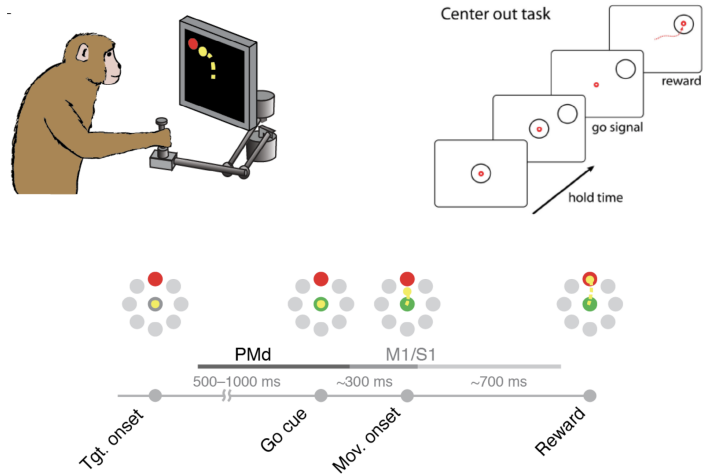}
\centering
\caption{Instructed delay center-out reaching task}
\label{fig:center_out}
\end{figure}

\subsection{Target dependent population activity}

Neural activity in motor and premotor cortices is recorded using multi-electrode arrays (MEAs). Given data for successive reaches to different targets (Fig. 26), we ask whether the pattern of population activity can be decoded to identify the target of each reach. We use a bin of size $\Delta = 300$ ms centered around the time of the go cue to characterize each reach by a point in an empirical neural space of dimension $N$, where $N$ is the number of recorded neurons. Each axis in this neural space represents the firing rate $f_{i}= y_{i} / \Delta$ of the $i$th neuron, where $y_i$ is the number of spikes emitted by neuron $i$ during the 300 ms window.

\begin{figure}[h]
\centering
\includegraphics[max width=200pt]{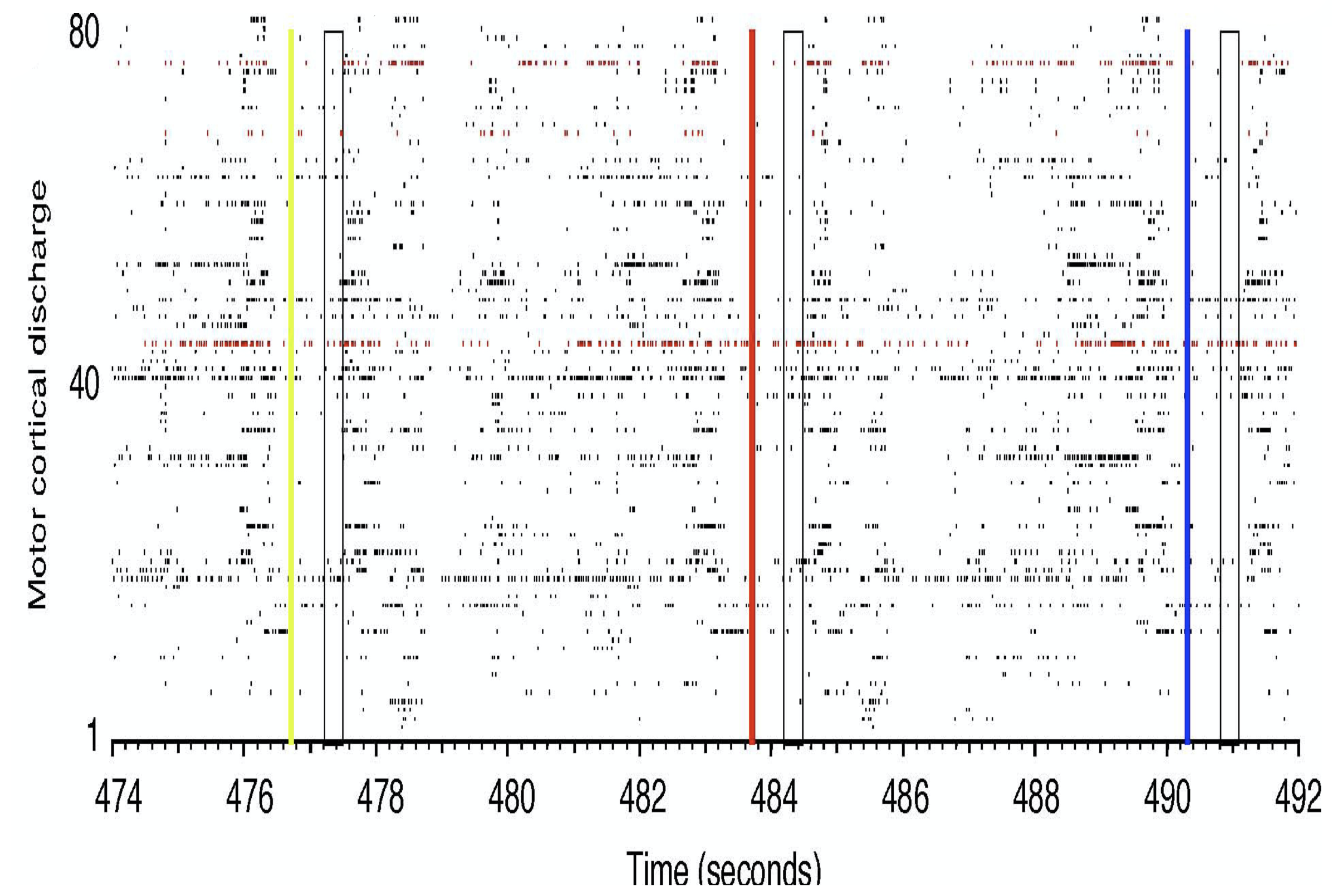}
\centering
\caption{Activity of $N = 80$ neurons during reaches to three targets. Yellow, red, and blue vertical lines indicate target presentation. Black vertical boxes show 300 ms windows centered around the corresponding go cue. }
\label{fig:center_out}
\end{figure}

In order to display the population activity, we choose three neurons (46, 70, and 78; rows shown in red in Fig. 26) and display the resulting three-dimensional neural space. Each point represents a trial, color coded by target (Fig. 27). Given this neural activity, is it possible to infer which target the monkey is going to reach? The answer seems to be NO, as the neural space shows no clustering, no organization according to target! 
\begin{figure*}[h]
\centering
\includegraphics[max width=300pt]{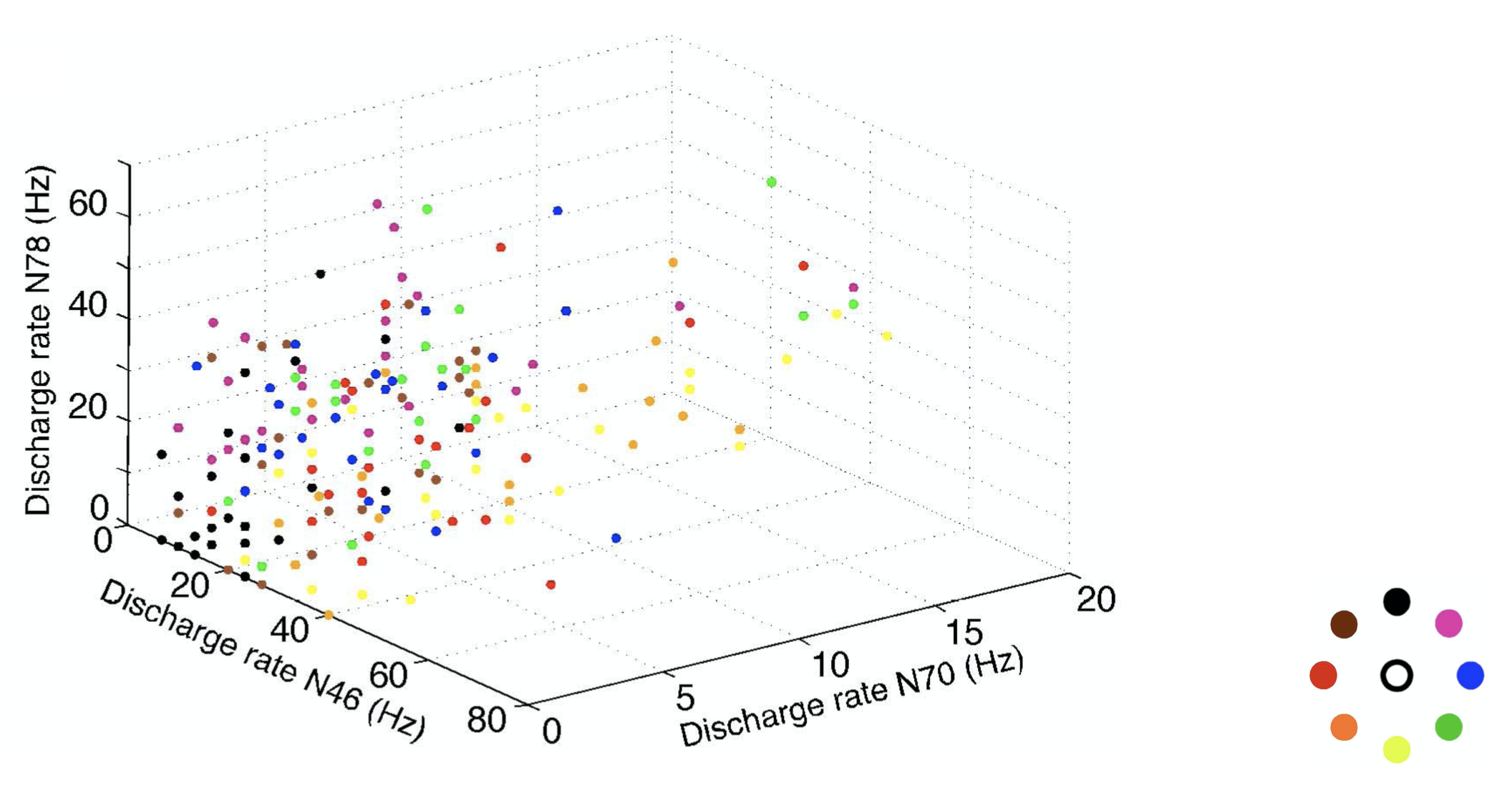}
\centering
\caption{Activity of the three selected neurons from trial to trial.}
\label{fig:activity-3-neurons}
\end{figure*}

This failure suggests that the data should be looked at from a different perspective. Let's recall the simple model of neurons tuned to the direction of motion \cite{georgopoulos1982relations}. In this model, a reach of extent $r$ in the direction $\theta$ would correspond to an activity 
\begin{equation}
f_{i} (r, \theta) =b_{i}+r a_{i}(1 / 2)\left[1+\cos \left(\theta-\theta_{i}\right)\right]~,
\end{equation}
where $b_{i}$ is the background activity, $a_{i}$ is the amplitude of activity modulation, and $\theta_{i}$ is the preferred direction of neuron $i$ (Fig. 21).
For a population of $N$ neurons, this model would predict
\begin{equation}
\vec{f}(r, \theta) =\vec{b}+(1 / 2) \ r \left[ \vec{a} + \cos (\theta) \ \vec {\phi_x} + \sin (\theta) \ \vec {\phi_y} \right]~,
\end{equation}
where 
\begin{equation}
\begin{split}
\vec{b} & =  (b_1, b_2, \dots, b_N) ~, \\
\vec{a} & =  (a_1, a_2, \dots, a_N) ~, \\
\vec{\phi_x} & = (a_1  \cos (\theta_1), a_2  \cos (\theta_2), \dots, a_N \cos (\theta_N)) ~, \\
\vec{\phi_y} & = (a_1 \sin (\theta_1), a_2 \sin (\theta_2), \dots, a_N \sin (\theta_N)) ~. \\
\end{split}
\end{equation}

A simulation of this model reveals that the data is organized in a low-dimensional ring structure: 
\begin{figure}[h]
\centering
\includegraphics[max width=200pt]{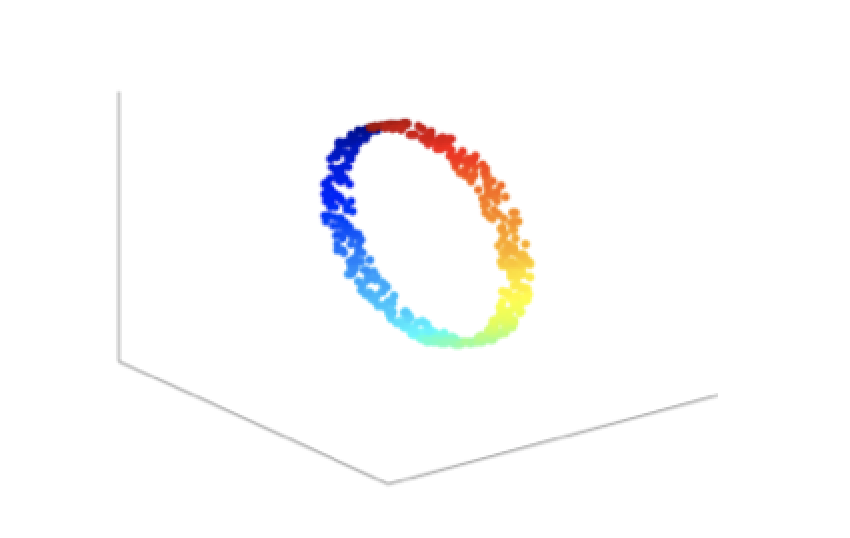}
\centering
\caption{Activity of $N = 100$ orientation selective neurons with preferred orientations uniformly distributed in $[0, 2 \pi]$ for planar reaches of extent $10 \leq r \leq 11$.}
\label{fig:ring}
\end{figure}

\noindent This observation suggests the use of standard dimensionality reduction methods to find this low-dimensional structure within the $N$-dimensional neural space.

\subsection{Linear dimensionality reduction: Principal components analysis (PCA)}

Consider data in the form of $N$-dimensional vectors. 
\begin{align}
\vec{x} = \left(\begin{array}{c} x_1 \\ x_2 \\ ... \\ x_N \end{array}\right) 
\end{align}
The data here is the $N$-dimensional vector of firing rates associated with each reach. Data for $M$ reaches results in an $N \times M$ matrix 
\begin{align}
    X= (\vec{x}_1 | \vec{x}_2 | ... | \vec{x}_M) \   =\left[\begin{array}{cccc}
x_{11} & x_{12} & \cdots & x_{1 M} \\
x_{21} & x_{22} & \cdots & x_{2 M} \\
\vdots & \vdots & & \vdots \\
x_{N 1} & x_{N 2} & \cdots & x_{N M}
\end{array}\right]
\end{align}

Estimate the mean firing rate of each neuron:
\begin{equation}
\hat{\mu}_{i}=\frac{1}{M} \sum_{k=1}^{M} x_{i k} \quad 1 \leq i \leq N
\end{equation}
and subtract it from the corresponding row (mean-centered data). Next, use the mean-centered data matrix $X$ to estimate the covariance of the data:
\begin{align}
&\hat{C}=\frac{1}{(M-1)} X X^{T} \\
&\hat{C}_{i j}=\frac{1}{(M-1)} \sum_{k=1}^{M}\left(x_{i k}-\hat{\mu}_{i}\right)\left(x_{j k}-\hat{\mu}_{j}\right) \ \ \text{for all neurons} \ \ 1 \leq i, j \leq N
\end{align}

The diagonalization of the covariance matrix yields eigenvectors and eigenvalues, the principal components:
\begin{align}
\widehat{C} \vec{u}_{v} = \lambda_{v} \vec{u}_{v} \ \ \text{for all} \ \ 1 \leq v \leq N
\end{align}

\subsubsection{Relation to singular value decomposition}
Consider the singular value decomposition (SVD) of the mean-centered data matrix $X$ :
\begin{equation}
X=U \Sigma V^{T}
\end{equation}

The columns of the $N \times N$ orthonormal matrix $U$ provide a basis for the neural space. The columns of the $M \times M$ orthonormal matrix $V$ provide a basis for the space of samples. 
Assume $M>N$. The $N \times M$ matrix $\Sigma$ consists of an $N \times N$ diagonal block and a rectangular block of zeros of size $N \times(M-N)$ to the right of the diagonal block.
The matrix $X$ has at most $\operatorname{rank} N$; only the leading $N$ columns of $V$, or $N$ rows of $V^{T}$ contribute to $X$.

Given the singular value decomposition of the data matrix $X$, the eigenvalue decomposition of $X X^T$ is given by
\begin{equation}
\begin{split}
X X^{T}&=\left(U \Sigma V^{T}\right)\left(V \Sigma^{T} U^{T}\right) \\
&=U\left(\Sigma \Sigma^{T}\right) U^{T}~.
\end{split}
\end{equation}
The empirical covariance follows:
\begin{equation}
\begin{split}
 \hat{C} &=\frac{1}{(M-1)} X X^{T}=U \Lambda U^{T} \\
 \Lambda &=\frac{1}{(M-1)} \Sigma \Sigma^{T} ~.
\end{split}
\end{equation}

\subsubsection{Dimensionality reduction}
Consider the diagonal matrix
\begin{equation}
\begin{split}
\Lambda &=\frac{1}{(M-1)} \Sigma \Sigma^{T} \\
\Lambda &=\left[\begin{array}{ccccc}\lambda_{1} & \cdots & 0 & \cdots & 0 \\ 0 & \cdots & \lambda_{K} & \cdots & 0 \\ 0 & \cdots & 0 & \cdots & \lambda_{N}\end{array}\right] \quad \ \text{with} \ \lambda_{1} \geq \cdots \geq \lambda_{K} \geq \cdots \geq \lambda_{N} ~.
\end{split}
\end{equation}

\begin{figure*}[h]
\includegraphics[max width=120pt]{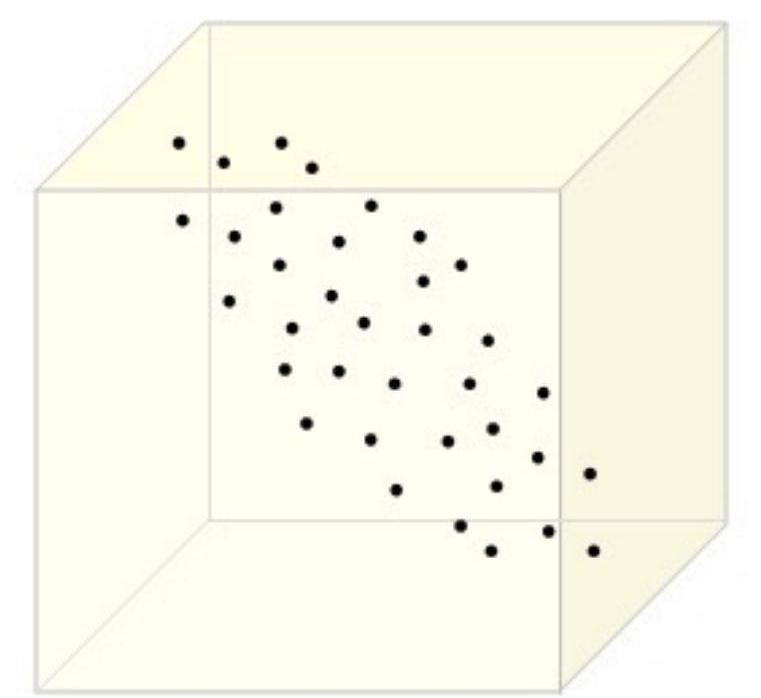}
\includegraphics[max width=120pt]{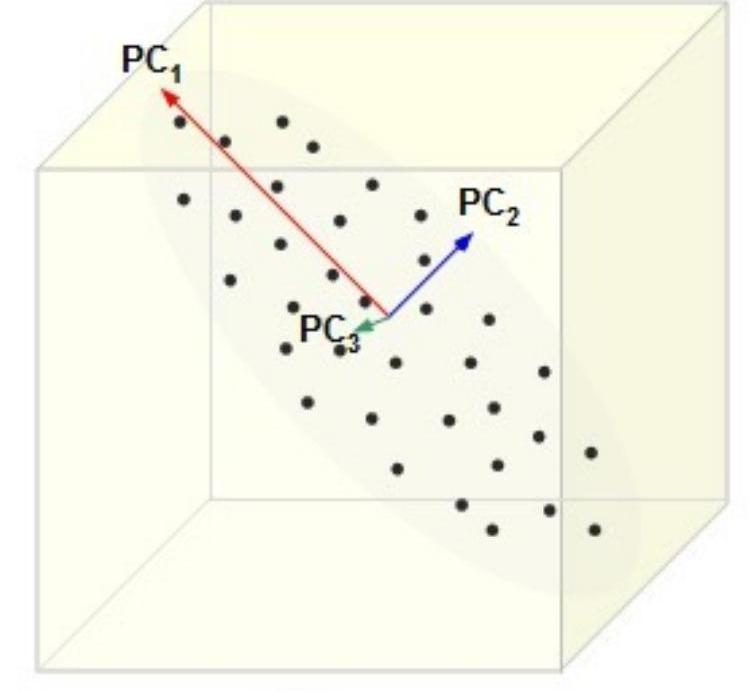}
\includegraphics[max width=120pt]{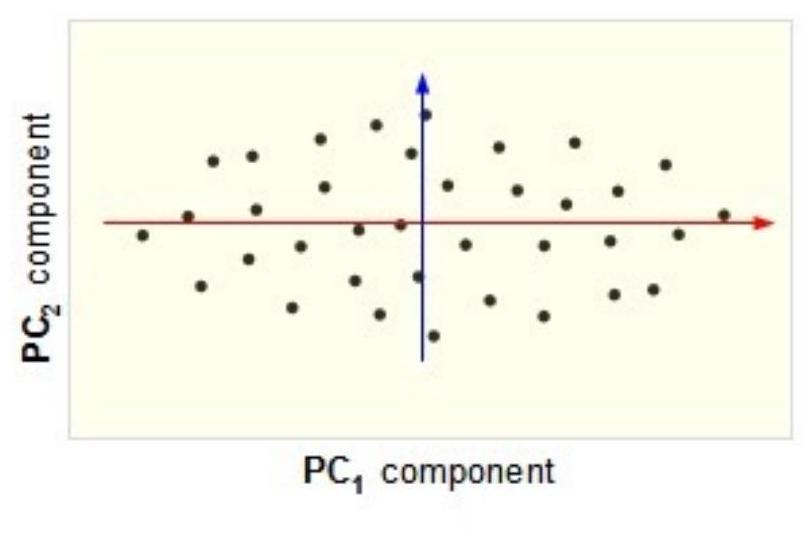}
\text{\hspace{1.5 cm} \textbf{(a)} \hspace{5 cm} \textbf{(b)} \hspace{5.5 cm} \textbf{(c)} \hspace{1 cm}}
\caption{Dimensionality reduction from a three dimensional original space \textbf{(a)} to a two dimensional space spanned by the leading eigenvectors \textbf{(b-c)}.}
\label{fig:dimensionality-red}
\end{figure*}

We keep the $K$ leading eigenvalues, $K<N$ and use the spectral decomposition of the covariance matrix to implement dimensionality reduction (Fig.~\ref{fig:dimensionality-red}): 
\begin{equation}
\hat{C} = \sum_{\mu=1}^{N} \lambda_{\mu} \vec{u}_{\mu} \vec{u}_{\mu}^{T} \quad \Rightarrow \quad \hat{C}=\sum_{\mu=1}^{K} \lambda_{\mu} \vec{u}_{\mu} \vec{u}_{\mu}^{T}
\end{equation}
The space of reduced dimensionality is spanned by the $K$ leading eigenvectors, as described by a matrix $U$ that has the leading eigenvectors $\vec{u}_{v}$, $1 \leq \nu \leq K$ as columns.
Then $\hat{C}=U \Lambda U^{T}$, where $\Lambda$ is the diagonal matrix of $K$ eigenvalues. 
The coordinates of the data points expressed in the new coordinate system are $Y=U^{T} {X}$. 

\subsubsection{Principal components as latent variables}
\begin{figure*}
\centering
\includegraphics[max width=100pt,scale=1.1]{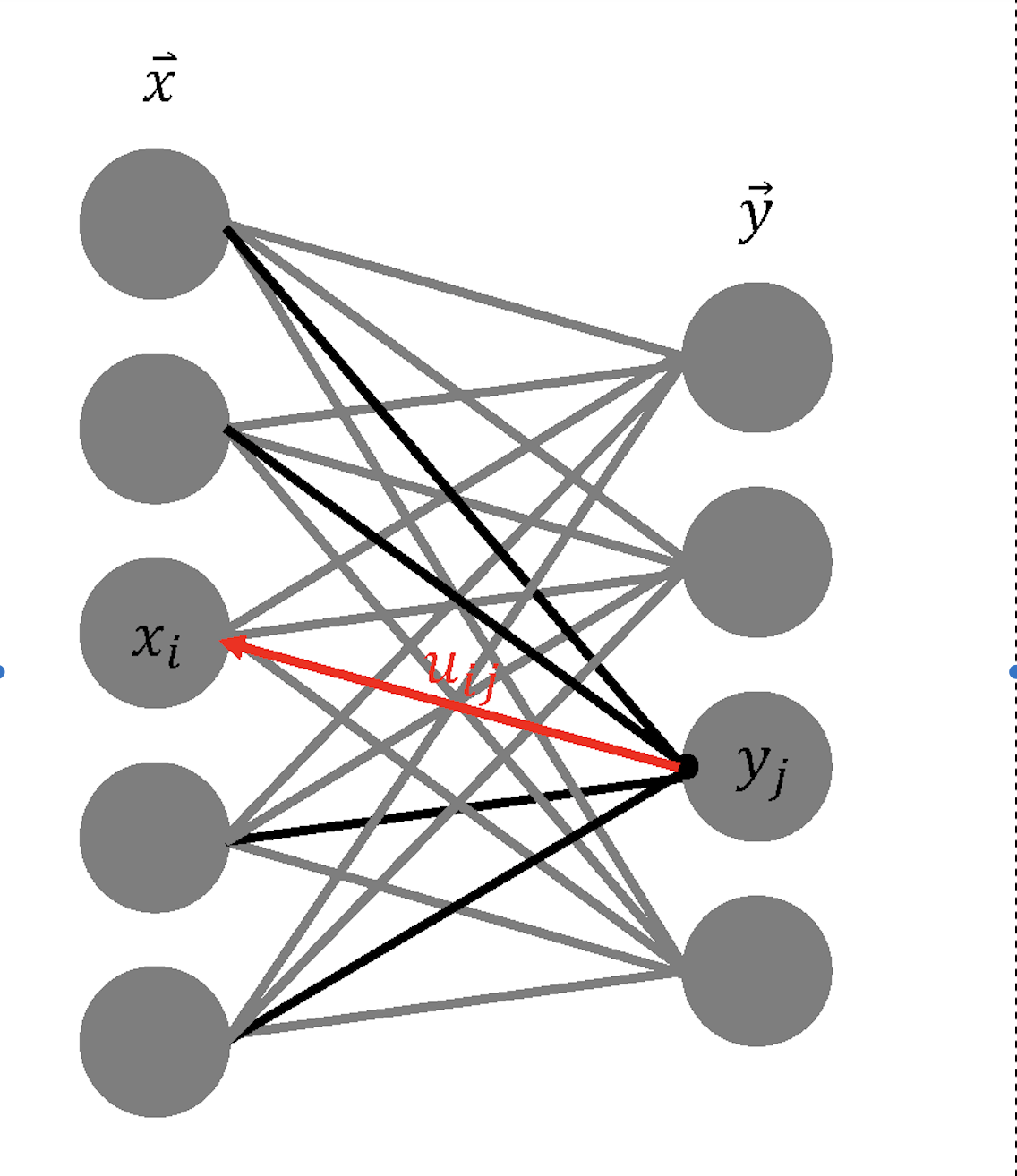}
\centering
\caption{Principal components as latent variables in a generative model.}
\label{fig:principal_as_latent}
\end{figure*}

We have described how to obtain $Y$ form $X$; similarly, we can reconstruct $X$ from $Y$ as follows (see Fig.~\ref{fig:principal_as_latent})
\begin{equation}
    X=UY \quad \Rightarrow \quad x_{i}=\sum_{j=1}^{K} u_{i j} y_{j}
\end{equation}
The $i$-th component $u_{i j}$ of the $j$-th eigenvector is the "weight" from $y_{j}$ to $x_{i}$. We can interpret this algorithm as a generative model in which the full-dimension data matrix $X$ is constructed from the low-dimensional data matrix $Y$.  
\begin{equation}
 P\left(y_{j}\right)=\mathcal{N}\left(0, \lambda_{j}\right) 
\quad \to \quad x_i = \sum_{j=1}^d u_{ij} y_j 
\end{equation}

Two extensions of PCA allow for the inclusion of noise in the generative model. Probabilistic PCA (PPCA) introduces homogeneous noise into the generative process: 
\begin{equation}
\begin{split}
x_i &=\sum_{j=1}^{d} u_{i j} y_{j}+\eta_{i} \\
P\left(\eta_{i}\right)&=\mathcal{N}\left(0, \sigma^{2}\right) \\
P\left(y_{j}\right) &=\mathcal{N}\left(0, \lambda_{j}\right) \ \\
P\left(x_i \mid y_j\right)&=\mathcal{N}\left(u_{i j} y_{j}, \sigma^{2}\right) \\
P\left(x_i \mid \vec{y}\right)&=\mathcal{N}\left(\sum_{j} u_{i j} y_{j}, \sigma^{2}\right)
\end{split}
\end{equation}
\begin{figure*}[h]
\centering
\includegraphics[max width=200pt]{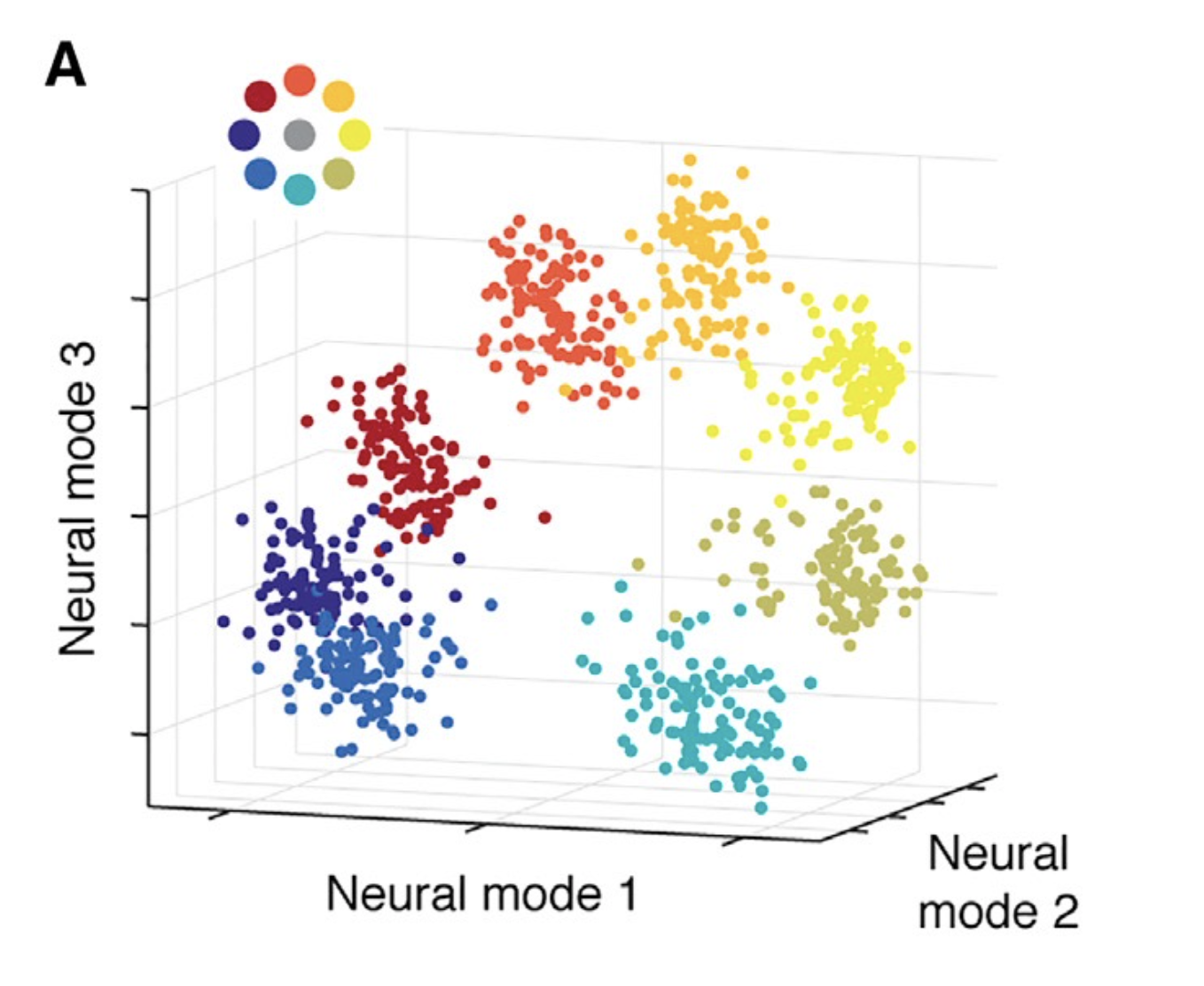} 
\centering
\caption{Neural activity in PMd preceding reaches in a center-out task. The three-dimensional representation is a projection onto a subspace spanned by the three leading principal components (neural modes).}
\label{fig:FAforPMd}
\end{figure*}

The constraint of homogeneous noise can be relaxed to allow for inhomogeneous noise whose amplitude depends on $i$. This leads from PPCA to Factors Analysis (FA): 
\begin{equation}
\begin{split}
P\left(\eta_{i}\right)&=\mathcal{N}\left(0, \sigma_i^{2}\right) \ \\
P\left(x_i \mid y_j\right)&=\mathcal{N}\left(u_{i j} y_{j}, \sigma_i^{2}\right) \\
P\left(x_i \mid \vec{y}\right)&=\mathcal{N}\left(\sum_{j} u_{i j} y_{j}, \sigma_i^{2}\right)
\end{split}
\end{equation}
Factor Analysis has been applied to neural data recorded from dorsal premotor cortex (PMd) during the execution of a center-out task \cite{shenoy2009FA}. This is a  region where neural activity is known to
correlate with the endpoint of an upcoming reach,  including both
direction and extent. In contrast to the random three-dimensional representation of Fig. 27, the projection of the data onto a three-dimensional space spanned by the three leading principal components reveals a structure that reflects the separation and spatial organization of the targets, see Fig. \ref{fig:FAforPMd}. 

\subsection{Dimensionality reduction: Neural modes and latent variables}

\begin{figure*}[h]
\centering
\includegraphics[max width=350pt]{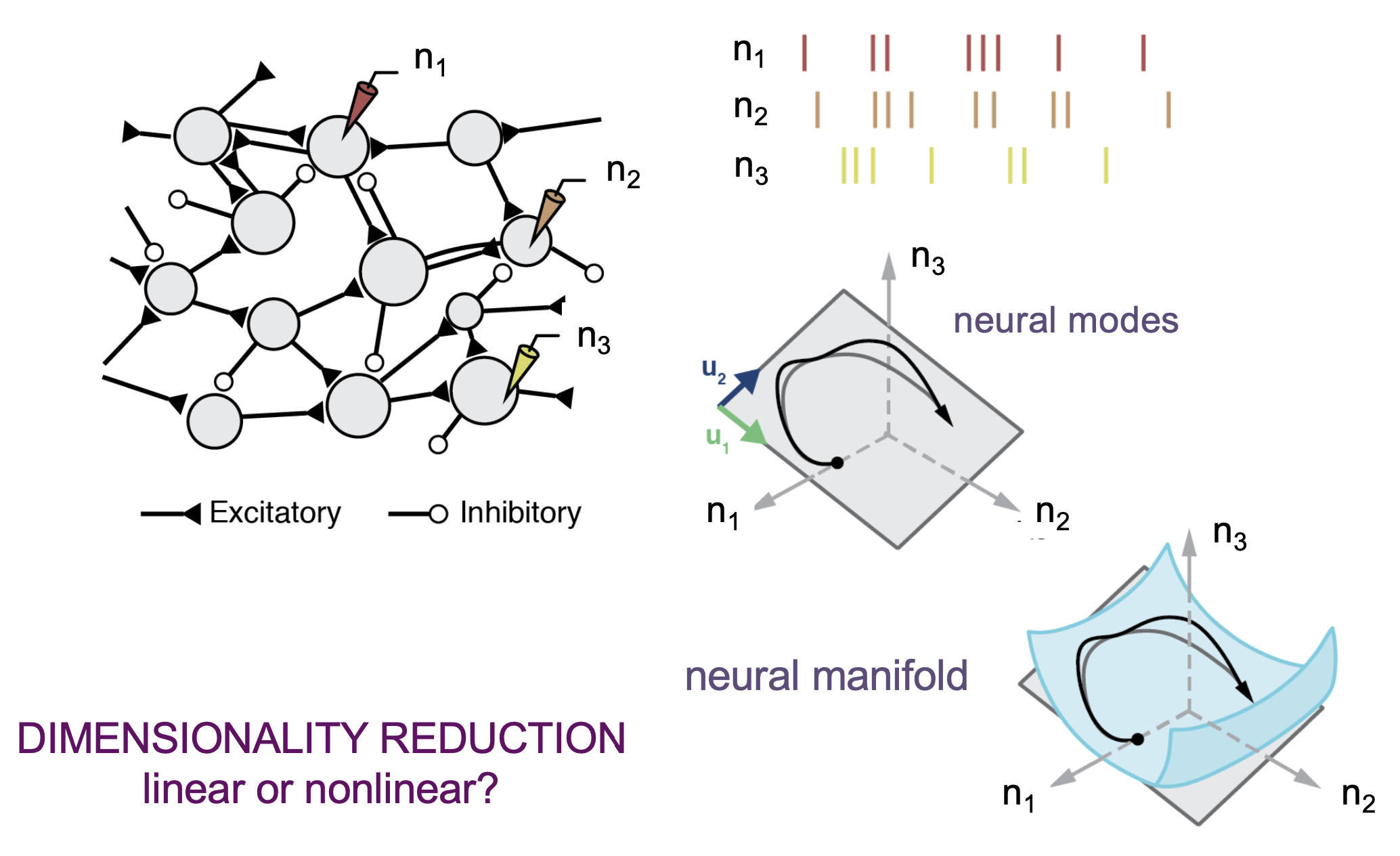}
\centering
\caption{Subsampling of the population neural activity defines an empirical neural space. The trajectory that describes the time evolution of the neural activity is confined to a low-dimensional manifold that can be either linear or nonlinear.}
\label{fig:neural-modes}
\end{figure*}

Let's consider a schematic scenario for the recording of neural population activity using multi-electrode arrays (Fig. \ref{fig:neural-modes}). The recording device subsamples the population activity and defines an empirical neural space of dimension $N$, the number of recorded neurons. Each axis in the empirical neural space corresponds to the firing rate of one of the recorded neurons. At every time bin of size $\Delta$, the population activity is described by a point in this empirical neural space. As time evolves, consecutive points define a trajectory that has been consistently found to be confined to a low-dimensional manifold within the empirical neural space \cite{yu2014dimred}. 
The observed low-dimensionality of population dynamics is not only restricted to the motor cortex, it is also observed in many other brain areas: frontal, prefrontal, parietal, visual, auditory, and olfactory cortices (see \cite{gallego2017} and references therein). A two-dimensional neural manifold that encodes for both position and accumulation of evidence has recently been characterized in hippocampus \cite{brody2021tank}.

\begin{figure*}[h]
\centering
\includegraphics[max width=300pt]{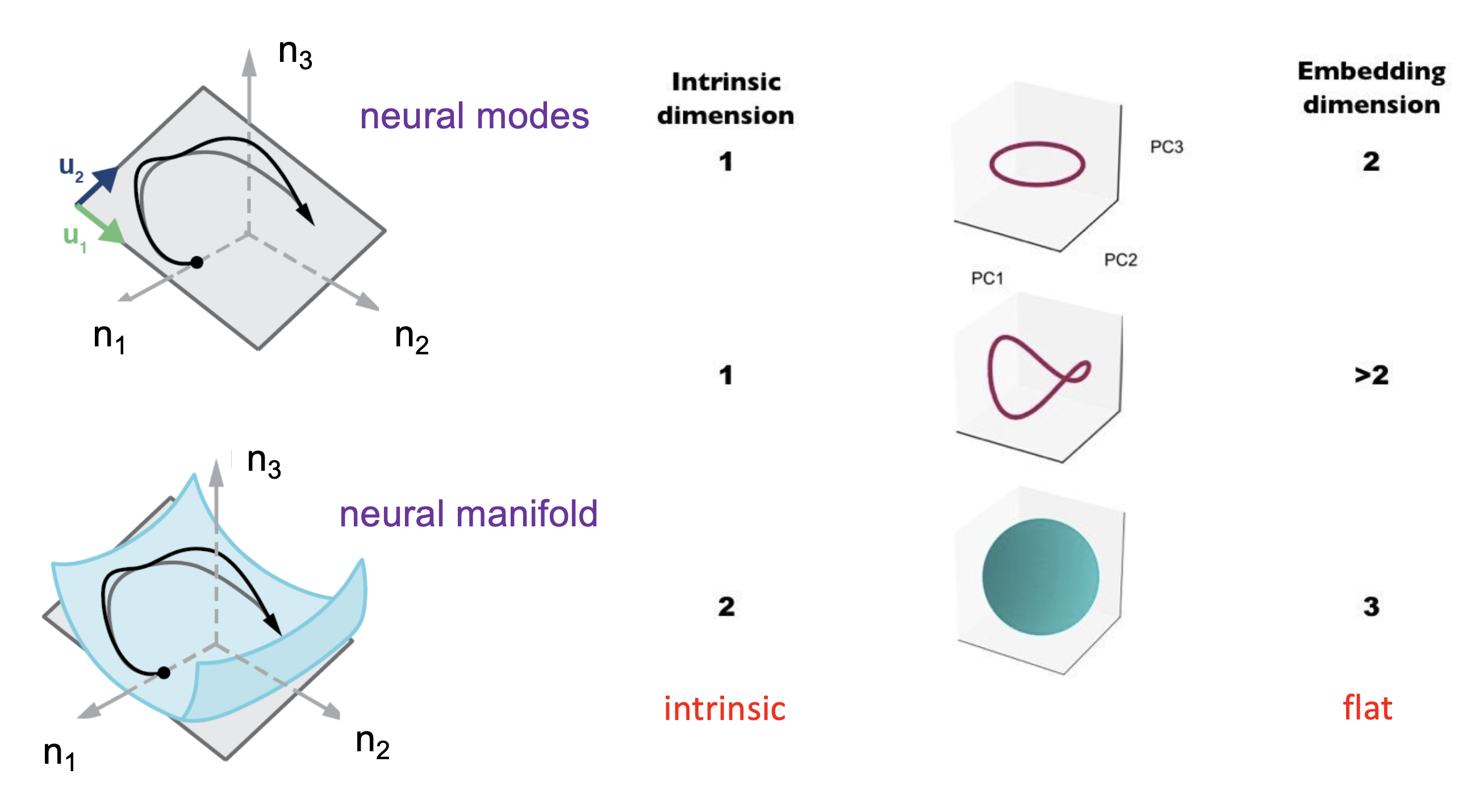}
\centering
\caption{Intrinsic vs embedding dimension.} 
\label{fig:intrinsic-vs-embedding}
\end{figure*}

An important aspect of these neural manifolds is whether they are linear or nonlinear \cite{altan2021plos}, an issue that leads to the concepts of {\it intrinsic dimension} and {\it embedding dimension} \cite{jazayeri2021interpreting}. As illustrated in Fig. \ref{fig:intrinsic-vs-embedding}, the intrinsic dimension is the true dimension of a nonlinear manifold, while the embedding dimension is the dimensionality of the smallest flat space that fully contains the manifold. The difference between these two dimensions is a proxy for the degree of manifold nonlinearity \cite{altan2021plos}. A useful approach to manifold identification is to use PCA to obtain a flat space whose dimensionality is low but not too low; this dimensionality must be large enough for the flat space to play the role of an embedding space. More sphisticated methods for nonlinear dimensionality reduction can then be used to identify the nonlinear manifold within the flat embedding space.

\subsection{Nonlinear dimensionality reduction: Isomap and multidimensional scaling}

\begin{figure*}[h]
\centering 
\includegraphics[max width=150pt]{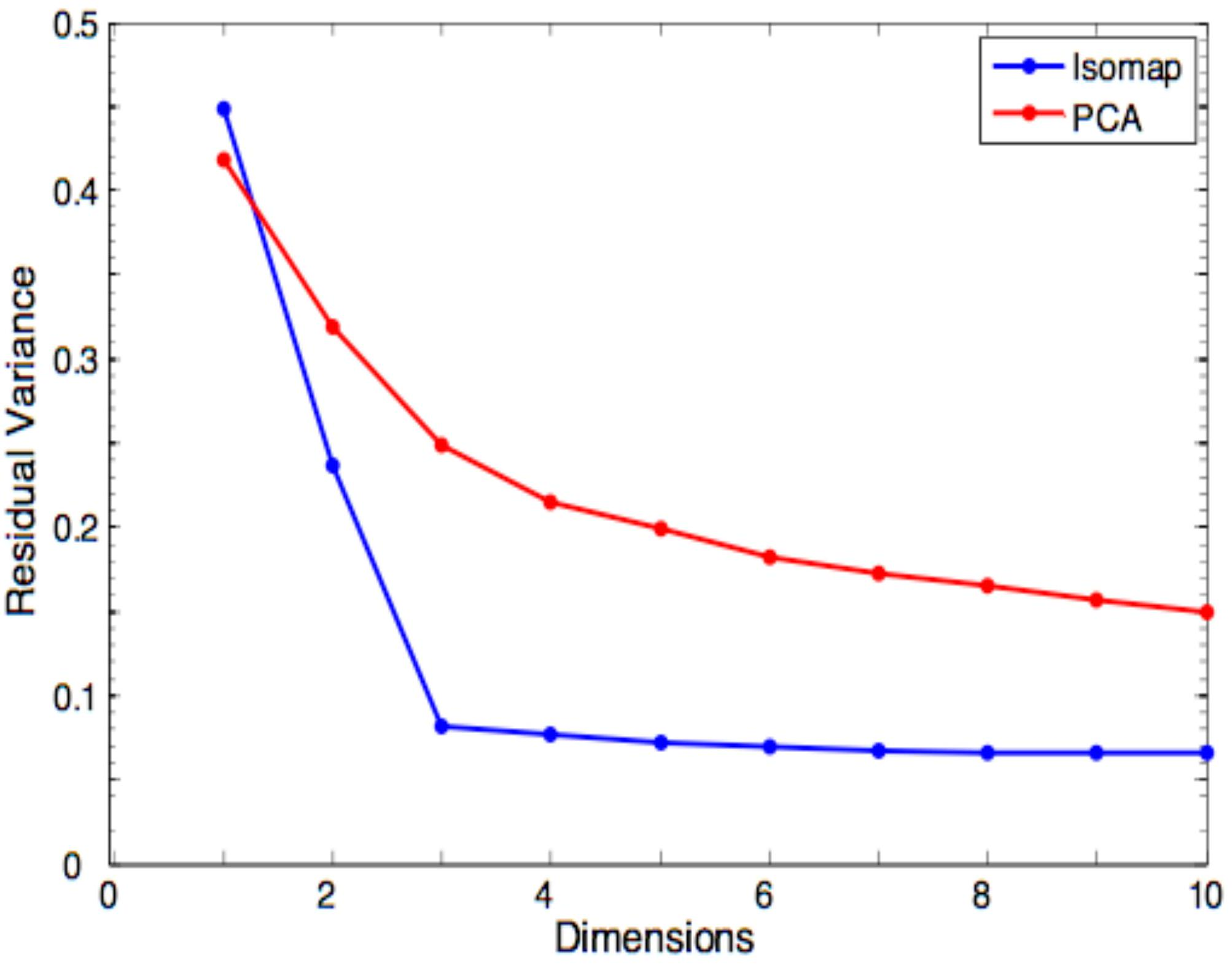}
\centering 
\caption{Eigenvalue spectrum for center-out task data obtained by PCA (red) and Isomap (blue).}
\label{fig:eigenvalues}
\end{figure*}

Let's go back to the center-out task that we introduced at the beginning of this lecture. 
If we apply PCA to data such as shown in Fig. 26, we find eigenvalues that gradually decrease as the dimensionality $K$ of the embedding space increases (Fig. \ref{fig:eigenvalues}, red curve). This gradual decay does not obviously define a threshold that identifies a specific value for $K$. 
In contrast, if we apply a nonlinear method for dimensionality reduction such as Isomap \cite{tenenbaum2000global} to the same data (Fig. \ref{fig:eigenvalues}, blue curve) we find that the first two eigenvalues dominate the others, which stay at an almost constant level. The magnitude of the eigenvalues $\lambda_i$ for $i \geq 3$ corresponds to the amplitude of the noise fluctuations that give the data cloud thickness in all nonsignificant dimensions. The leading dimensions describe a nonlinear manifold of intrinsic dimension two. 

\begin{figure*}[h]
\centering 
\includegraphics[max width=250pt]{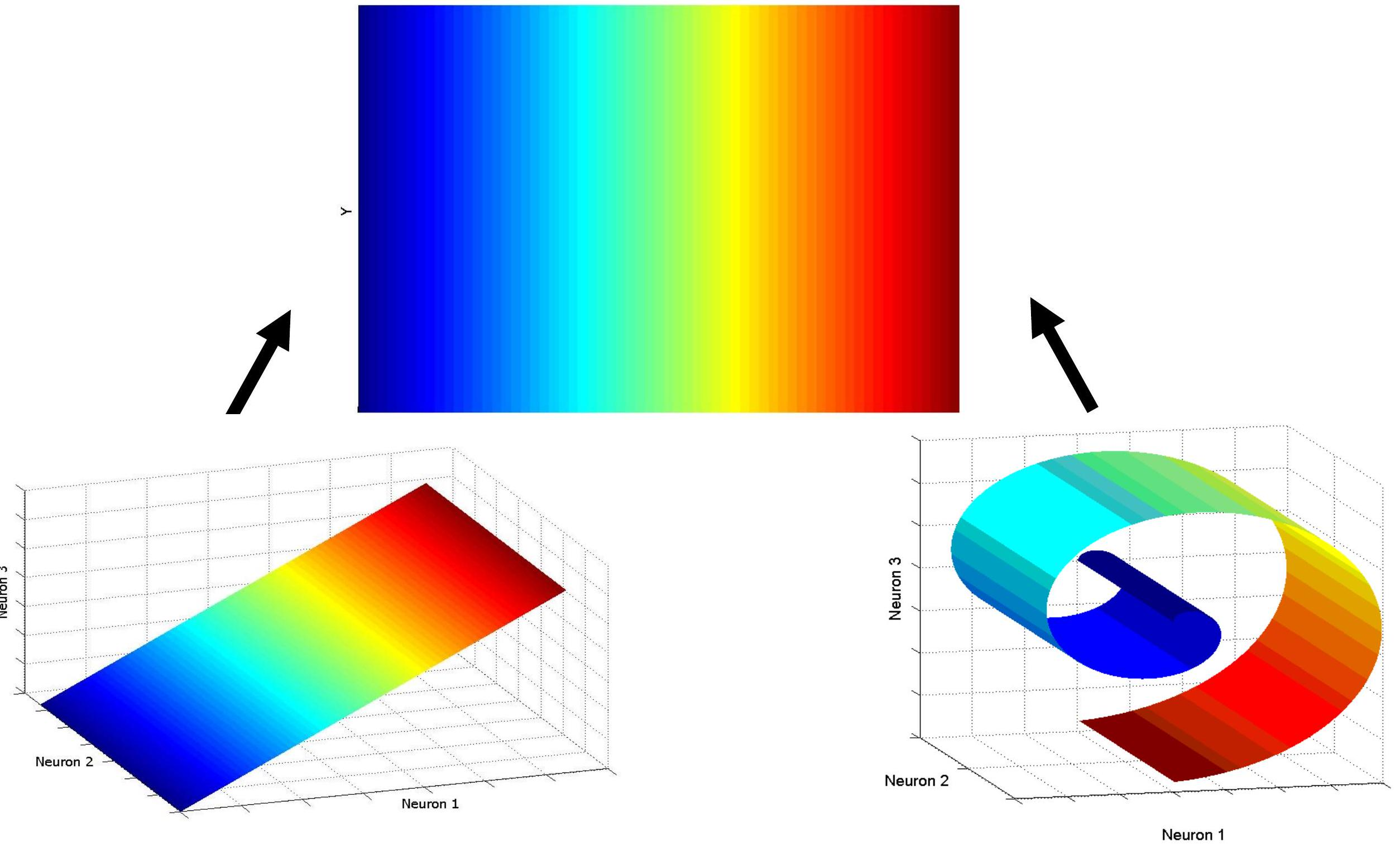}
\centering 
\caption{Isomap flattens curved manifolds.}
\label{fig:manifold-flattening}
\end{figure*}

\begin{figure*}[h]
\centering 
\includegraphics[max width=300pt]{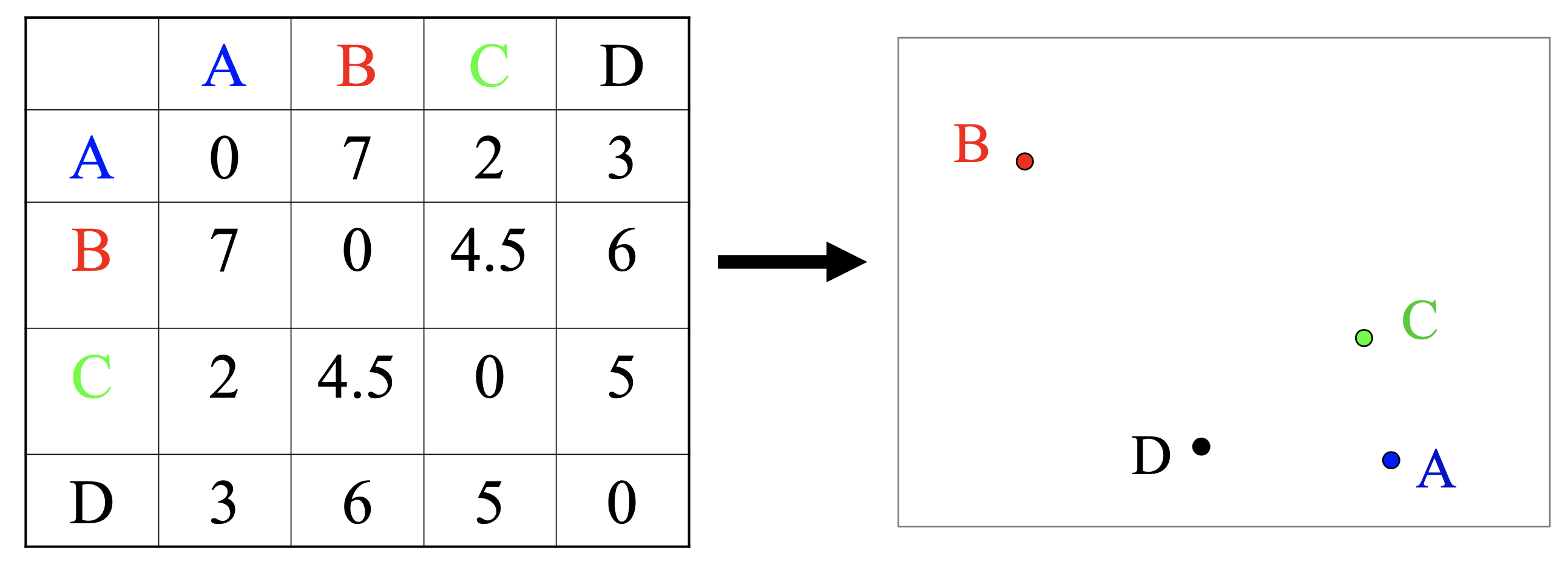}
\centering 
\caption{Matrix of dissimilarities between $M=4$ objects (left); representation of these objects as points in an Euclidean space  (right).}
\label{fig:MDS}
\end{figure*}

How does Isomap work \cite{tenenbaum2000global}? It flattens the curved manifold into a flat manifold of the same dimension (Fig. \ref{fig:manifold-flattening}).  
Isomap is based on MultiDimensional Scaling (MDS), a method devised to find a solution to the following problem (Fig. \ref{fig:MDS}): given an $M \times M$  matrix of dissimilarities between $M$ objects, is it possible to represent these objects as $M$ points in an Euclidean space such that the matrix of Euclidean distances between them matches as well as possible the original dissimilarity matrix \cite{cox2000mds}? 

Consider data in the form of $N$-dimensional vectors, such as an $N$-dimensional vector of firing rates associated with each reach. Data for $M$ reaches result in an $ N \times M$ data matrix $X$:
\begin{equation}
\begin{array}{r}
\vec{x}=\left(\begin{array}{c}
x_{1} \\
x_{2} \\
\cdots \\
x_{N}
\end{array}\right) \quad  \Longrightarrow   \quad 
X = \left(\vec{x}_{1}\left|\vec{x_{2}}\right| \ldots \mid \vec{x}_{M}\right)
\end{array}
\end{equation}
If the matrix $X$ is hidden from us, but we are given instead an $M \times M$ matrix $S$ of squared distances between the points, can we reconstruct the matrix $X$?
Let's start by considering the case where the original distances are Euclidean:
\begin{equation}
S_{i j}=d_{i j}^{2}=\left(\vec{x}_{i}-\vec{x}_{j}\right)^{T}\left(\vec{x}_{i}-\vec{x}_{j}\right)~.
\end{equation}
The scalar product between data points can be written as:
\begin{equation}
\vec{x}_{i}^{T} \vec{x}_{j}=-\ (1 / 2) \ \left(S_{i j}-\|\vec{x}_{i}\|^{2}-\|\vec{x}_{j}\|^{2}\right)~.
\end{equation}
In matrix form,
\begin{equation}
\vec{x}_{i}^T \vec{x}_{j}=\left(X^{T} X\right)_{i j}~,
\end{equation}
and 
\begin{equation}
S_{i j}-\|\vec{x}_{i}\|^{2}-\|\vec{x}_{j}\|^{2}=(J S J)_{i j}~, 
\end{equation}
where $J$ is the $M \times M$ centering matrix $J=I-(1 / M) \ e e^{T}$ with $e$ being a column vector with all entries equal to $1$. Then
\begin{equation}
X^T X =- \ (1 / 2) \  J S J~.
\end{equation}
From this equation the data matrix $X$ can be easily obtained:
\begin{equation}
X^{T} X=U \Lambda U^{T} \Longrightarrow X=\Lambda^{1 / 2} U^{T}~.
\end{equation}

If the distance matrix to which this calculation is applied is based on Euclidean distances, this process allows us to recover the data matrix $X$ from the matrix $S$ of squared distances. A reduction of the dimensionality of the Euclidean representation follows from truncating of the number of eigenvalues in the diagonal matrix $\Lambda$ from $M$ to $K$; this restricts to $K$ the number of eigenvectors used to reconstruct $X$.
It can be proved that this truncation is equivalent to PCA, which is based on the diagonalization of $X X^{T}$.

When applied to an arbitrary matrix $S$ of squared distances, the method follows similar steps.  An  `inner product' matrix $Y$ is defined through the centering operation: $Y=- \ (1 / 2) \ J S J$, followed by the diagonalization of $Y: Y=U \Lambda U^{T}$ and the identification of the data matrix $X$ as $X=\Lambda^{1 / 2} U^{T}$.
This procedure minimizes a cost function $E$ that measures the Frobenius norm of the difference between two matrices: the matrix $Y$ defined by the original $S$ and the inner product matrix $X^{T} X$ obtained from the Euclidean representation of the data:
\begin{equation}
E(X)=\left\|X^{T} X-Y\right\|_{F}~.
\end{equation}

\begin{figure*}[h]
\centering
\includegraphics[max width=300pt]{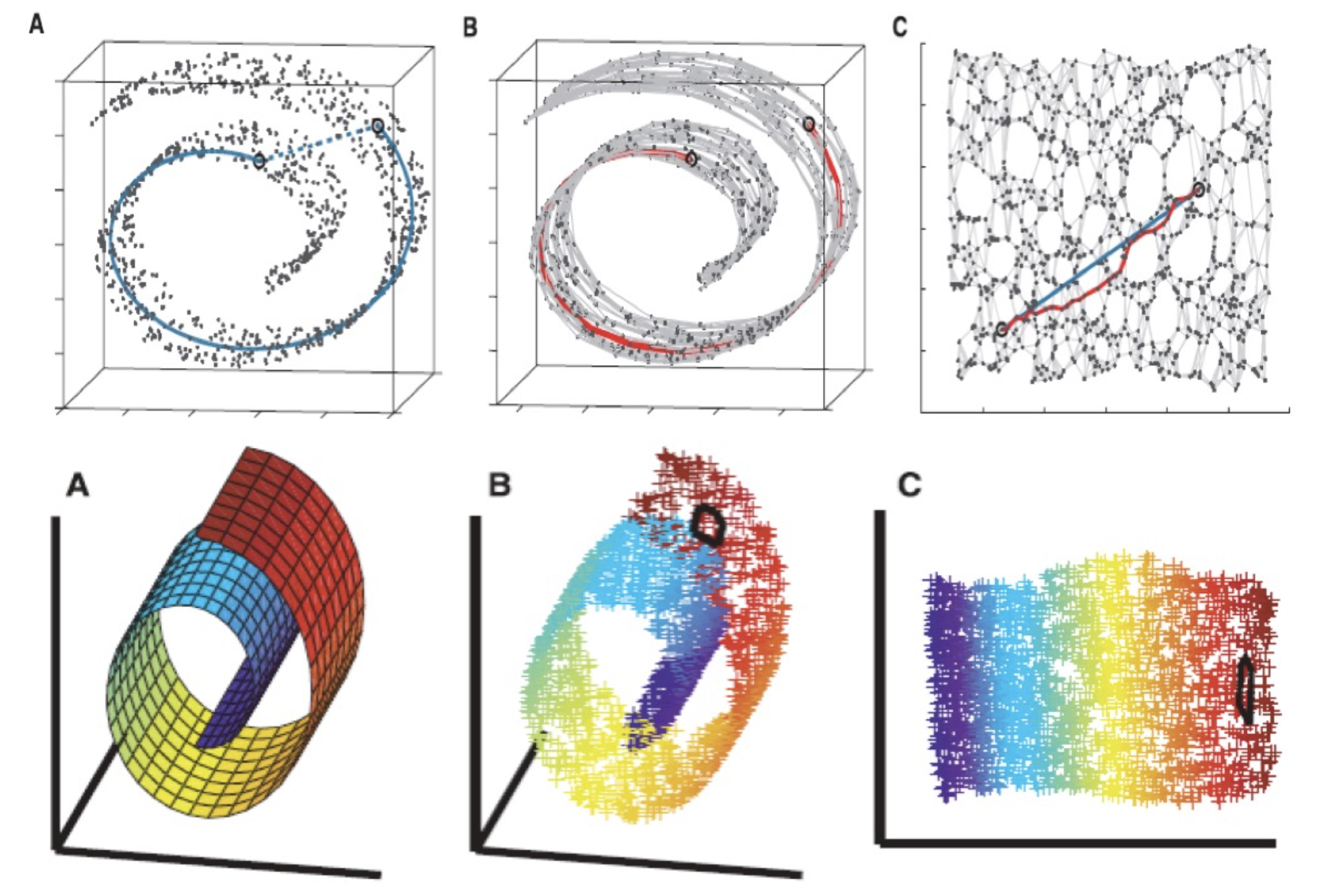} 
\centering
\caption{The distance between two points that are not nearest neighbors is defined through an empirical Geodesic: the shortest path that connects the two points through data points. This Geodesic distance will typically be larger than the Euclidean distance between the two points (see dashed blue line in A).}
\label{fig:geodesics}
\end{figure*}

The beauty of the Isomap approach is based on the proposed method for computing the matrix $S$ of square distances, which is obtained from the data by constructing empirical Geodesics \cite{tenenbaum2000global}. For each data point, a number $P$ of nearest neighbors is chosen, and the short distances from the chosen point to its $P$ neighbors are computed as Euclidean. The distance between two points that are not nearest neighbors follows from walking in the manifold. Every step in this path involves a known short Euclidean distance between neighboring points; the shortest path that connects the two end points defines their Geodesic distance (Fig. \ref{fig:geodesics}, top row)~.

\begin{figure*}[h]
\centering
\includegraphics[max width=250pt]{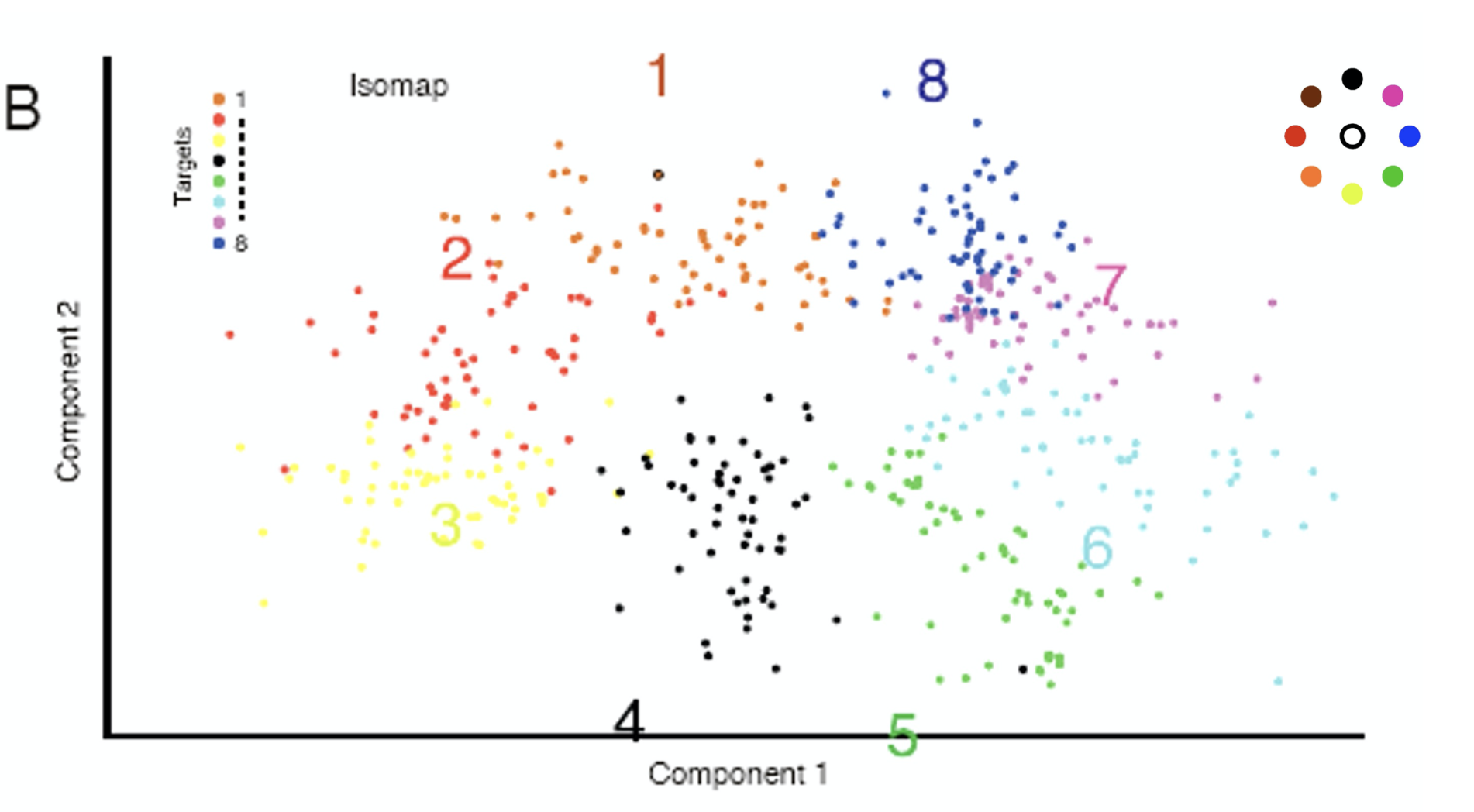}
\centering
\caption{Two dimensional representation of neural data for the center-out task shows clusters that capture the spatial organization of the targets.}
\label{fig:Isomap-targets}
\end{figure*}

\begin{figure*}[h]
\centering
\includegraphics[max width=150pt]{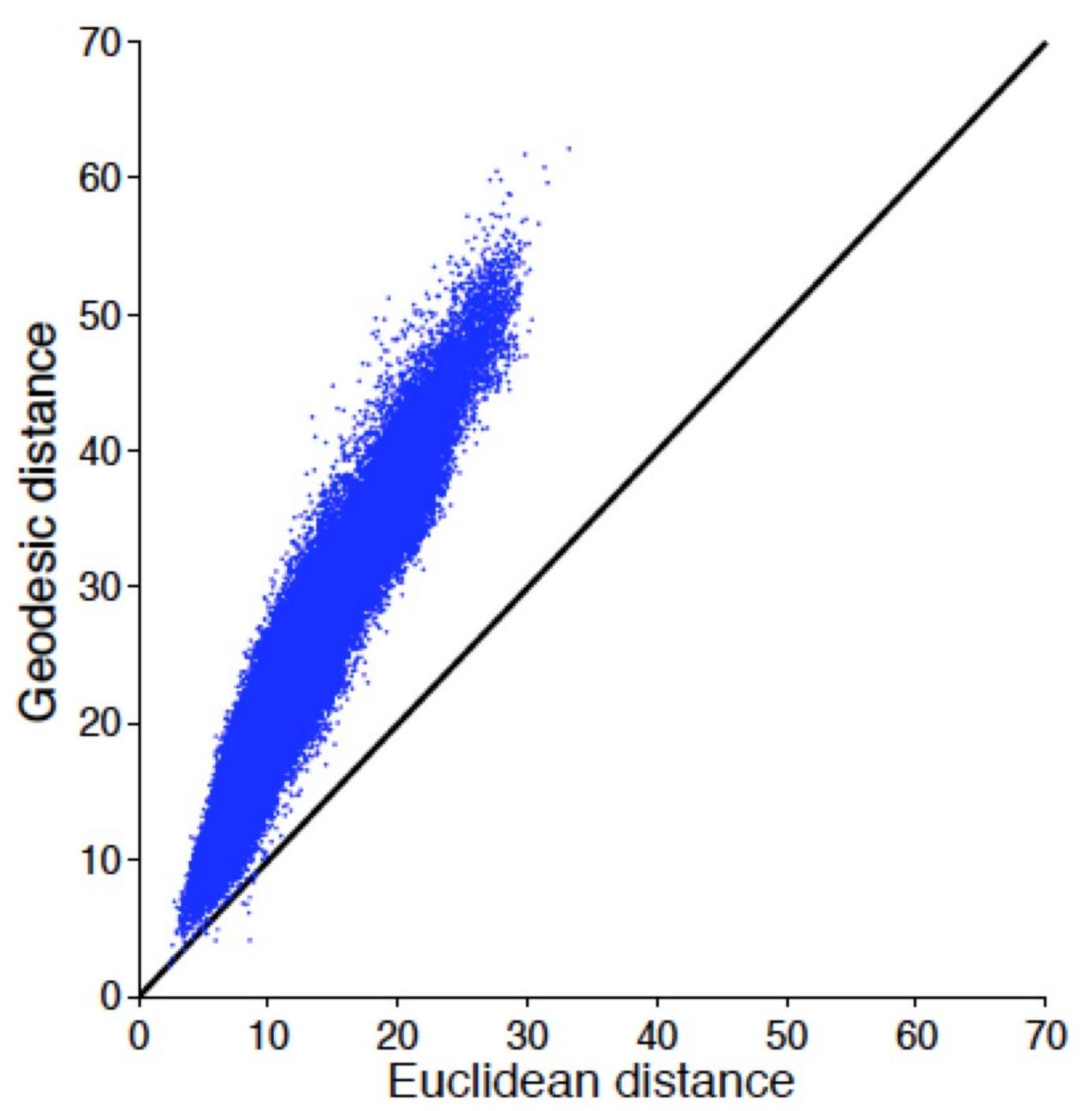}
\centering
\caption{Geodesic vs Euclidean distances.}
\label{fig:geodesic-euclidean}
\end{figure*}

When applied to the Swiss roll data  \cite{tenenbaum2000global}, Isomap results in a flat two-dimensional representation that preserves the neighborhood and distance relations among data points (Fig. \ref{fig:geodesics}, bottom row). When applied to neural data recorded during the execution of the center-out task, the Isomap eigenvalue spectrum (Fig. \ref{fig:eigenvalues}, blue curve) signals a two-dimensional curved manifold.  The Isomap projection then provides a flat two-dimensional clustered representation that captures the spatial organization of the targets (Fig. \ref{fig:Isomap-targets}).  
The degree of nonlinearity of the two-dimensional manifold can be inferred from a comparison of pairwise Geodesic and Euclidean distances (Fig. \ref{fig:geodesic-euclidean}). Note that the distinction between these two distances vanishes for the short distances between nearest neighbors, and increases monotonically for larger distances. 

\subsection{Summary}
\begin{itemize}
  \item The dynamics of neural population activity is typically confined to low-dimensional manifolds within a high-dimensional neural space. 

  \item These manifolds can be linear or nonlinear. A nonlinear manifold is  characterized by its {\it intrinsic dimension} and by the {\it embedding dimension} of the smallest flat space that fully contains it. The difference between these two dimensions is a proxy for the degree of manifold nonlinearity. 
  
  \item MultiDimensional Scaling in combination with the empirical Geodesic distances proposed by Isomap provide a useful tool for obtaing flat representations of nonlinear manifolds. 
  
\end{itemize}



\begin{acknowledgments}
These are notes from the lecture of Sara Solla given at the summer school "Statistical Physics \& Machine Learning", that took place in Les Houches School of Physics in France from 4th to 29th July 2022. The school was organized by Florent Krzakala and Lenka Zdeborová from EPFL.

\end{acknowledgments}


\bibliography{main.bib}


\end{document}